\documentclass[10pt,twocolumn,letterpaper]{article}
\usepackage[pagenumbers]{cvpr}

\usepackage{array,booktabs,makecell,multirow,tabularx}
\usepackage{caption,cuted,graphicx}
\usepackage[percent]{overpic}
\usepackage[table]{xcolor}

\newcolumntype{Y}{>{\centering\arraybackslash}X}
\newcommand{\para}[1]{\vspace{3pt}\noindent\textbf{#1}}

\definecolor{RFBlueRow}{RGB}{236,244,251}
\definecolor{RFGrayRow}{RGB}{245,245,245}
\colorlet{RFAltRow}{RFGrayRow}

\newcommand{\RFtablefont}{\footnotesize}
\newcommand{\RFwidetablefont}{\scriptsize}
\newcommand{\RFtablespacing}{\renewcommand{\arraystretch}{1.08}}
\newcommand{\RFwidetablespacing}{\renewcommand{\arraystretch}{1.06}}

\newcommand{\RFwidegroup}[1]{{\scriptsize #1}}

\newcommand{\RFhighlight}{\rowcolor{RFBlueRow}}
\newcommand{\RFbest}[1]{\textbf{#1}}
\newcommand{\RFmethod}[1]{\textbf{#1}}

\newcommand{\RFwideSub}[1]{\raisebox{-0.32ex}{\fontsize{5pt}{5pt}\selectfont #1}}

\newcommand{\RFwidehSSIM}{{\scriptsize SSIM\,\ensuremath{\uparrow}}}
\newcommand{\RFwidehSpec}{{\scriptsize MAE\RFwideSub{Spec}\,\ensuremath{\downarrow}}}
\newcommand{\RFwidehPwr}{{\scriptsize MAE\RFwideSub{Pwr}\,\ensuremath{\downarrow}}}

\newcommand{\RFhSSIM}{{\scriptsize SSIM\,\ensuremath{\uparrow}}}
\newcommand{\RFhSpec}{{\scriptsize MAE$_{\mathrm{Spec}}$\,\ensuremath{\downarrow}}}
\newcommand{\RFhPwr}{{\scriptsize MAE$_{\mathrm{Pwr}}$\,\ensuremath{\downarrow}}}

\newcommand{\appref}[1]{\hyperref[#1]{\textcolor{cvprblue}{Appendix~\ref*{#1}}}}

\definecolor{cvprblue}{rgb}{0.21,0.49,0.74}
\usepackage[pagebackref,breaklinks,colorlinks,allcolors=cvprblue]{hyperref}

\title{RFWeave: Generalizable Wireless Propagation Prediction \\ Across Scenes and Transceiver Configurations}

\author{%
Shurui Xu$^{1,\ast}$ \quad Xiao Ye$^{1,\ast}$ \quad Yijie Bian$^{2,\ast}$ \quad Junjie Ai$^{1,\ast}$ \quad Yuxuan Chen$^{1}$ \\
Weicong Chen$^{1}$ \quad Wei Guo$^{2}$ \quad Wankai Tang$^{1,\dagger}$ \quad Jun Zhang$^{2,\dagger}$ \quad Shi Jin$^{1,\dagger}$ \\
 {$^{1}$Southeast University \quad $^{2}$The Hong Kong University of Science and Technology }
}

\makeatletter

\patchcmd{\@maketitle}
  {\vspace*{24pt}}
  {\vspace*{14pt}}
  {}{}

\patchcmd{\@maketitle}
  {\vspace*{12pt}}
  {\vspace*{-22pt}}
  {}{}

\makeatother

\begin{document}
\maketitle

\begingroup
\renewcommand{\thefootnote}{\fnsymbol{footnote}}
\footnotetext[1]{Co-first authors.\quad
\textsuperscript{\ensuremath{\dagger}} Corresponding authors.
\quad \href{https://rfweave.github.io}{\texttt{\textbf{Project Page}}}.}
\endgroup

\begin{strip}
\centering
\includegraphics[width=0.98\linewidth]{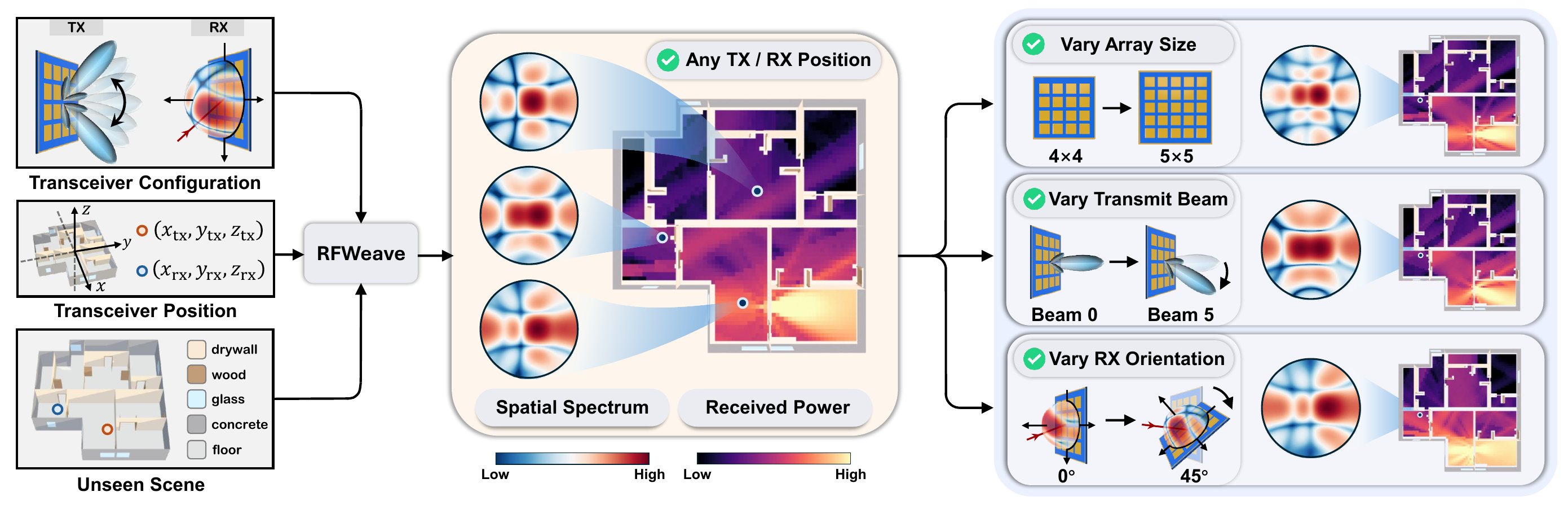}
\captionsetup{hypcap=false}
\captionof{figure}{\textbf{RFWeave} is a unified feedforward model for wireless propagation prediction. Given a material-annotated scene, queried transceiver positions, and a transceiver configuration, RFWeave predicts the receive-side spatial spectrum and received power in a single forward pass. It supports arbitrary transceiver position queries and generalizes across array sizes, receiver orientations, transmit beams, and their combinations on unseen scenes. By querying multiple receiver positions, RFWeave can also generate received power maps.}
\captionsetup{hypcap=true}\label{fig:rfweave_teaser}
\end{strip}

\begin{abstract}
Wireless propagation modeling is foundational to the evaluation and design of modern communication systems.
While recent learning-based methods have shown promising generalization across transceiver positions and unseen scenes, they remain largely confined to a fixed configuration of array size, receiver orientation, and transmit beam.
In this work, we introduce \textbf{RFWeave}, a unified feedforward model for wireless propagation prediction across scenes, transceiver positions, and transceiver configurations.
Its core innovation is a flexible antenna representation that accommodates varying array structures while capturing spatial relations between transmitter and receiver elements.
Our conditioning mechanism allows the model to use a shared Transformer backbone across diverse transceiver configurations, enabling joint prediction of received power and the angle-of-arrival spatial spectrum.
Trained on a large-scale simulated dataset spanning 10,000 indoor scenes and more than 5 million channel records, RFWeave achieves state-of-the-art performance across a wide range of generalization settings, improving spatial spectrum structural similarity by up to 15.6\% and reducing received power mean absolute error by up to 35.4\% over the strongest baselines on unseen scenes.
Additional experiments on measured DICHASUS data and in beam management further demonstrate RFWeave's broader applicability to real-world channels and downstream wireless tasks.
\end{abstract}
\vspace{-5pt}
\section{Introduction}\label{sec:intro}

For modern communication, accurate wireless propagation modeling is foundational to link-level evaluation and system design. It characterizes how radio-frequency (RF) signals propagate through physical environments and are observed at receivers. High-fidelity propagation prediction requires accounting for scene geometry and materials, transceiver positions, and transceiver configurations. While current learned propagation models can generalize across transceiver positions \citep{zhao2023nerf2,wen2025wrfgs}, they typically require scene-specific optimization and assume a fixed transceiver configuration, limiting direct reuse across changing environments.

In pursuit of generalizable wireless propagation prediction, prior work has largely followed two main directions, but each remains limited in scope. Cross-scene propagation methods learn to predict wireless propagation from 3D scene geometry and material properties \citep{hehn2025wigatr,yang2026graf,chen2025rfscape}. However, they typically assume a common transceiver configuration, with array size, receiver orientation, and transmit beam held fixed. In parallel, transceiver configuration has mainly been studied in downstream tasks such as beam prediction, selection, and beamforming \citep{xu2020scene_beam,rezaie2022orientation_beam,rezaie2022multipanel,kilinc2026heterogeneity,bian2026gsbf}. These methods incorporate antenna orientation, array structure, and beam information to improve communication decisions. Yet these lines of work have largely treated scene and transceiver variation as separate problems. In practice, these factors often vary together: mobile transceivers continuously change position and orientation, transmit beams are dynamically reconfigured, and deployment environments vary across scenarios \citep{charan2022challenge}. This motivates a unified propagation model that generalizes jointly across scenes, transceiver positions, and transceiver configurations.

Beyond scene and position generalization, transceiver configuration changes introduce a distinct representation challenge. Array size changes both the number and spatial arrangement of antenna elements. Receiver orientation changes the geometric relationship between the array and its surroundings, whereas the transmit beam changes the directional excitation of the scene. These configuration factors may also vary jointly at test time. Consequently, changes in transceiver configuration affect not only the values of model inputs, but also the structure and geometry that the model must represent.

In this work, we introduce \textit{\textbf{RFWeave}}, a unified feedforward model for wireless propagation prediction across scenes, transceiver positions, and transceiver configurations, as illustrated in \cref{fig:rfweave_teaser}. To accommodate varying array structures, RFWeave represents transceivers with variable-length tokens that preserve antenna element structure across array sizes. The material-annotated scene point cloud is compressed into scene tokens that summarize local geometry and material properties. RFWeave further encodes the pairwise geometry between transmitter (TX) and receiver (RX) elements and aggregates these interactions into element-indexed propagation tokens. A shared Transformer backbone then refines the propagation tokens using the scene and transceiver representations. Because receiver orientation and transmit beam are introduced only at the prediction stage, the backbone representation can be reused across multiple configuration queries.

We construct a comprehensive wireless propagation dataset by applying far-field synthesis to multipath channel records from WiPTR \citep{hehn2025wigatr} and Wi3R \citep{orekondy2023winert}, yielding spatial spectrum and received power targets. The dataset contains 10,000 scenes and over 5 million channel records, while evaluation includes a wide range of unseen configurations. Extensive evaluations show that RFWeave improves spatial spectrum structural similarity (SSIM) by up to 15.6\% and reduces received power mean absolute error (MAE) by up to 35.4\% over the strongest baselines on unseen scenes. These gains persist under unseen array sizes, receiver orientations, and transmit beams, as well as joint configuration variation. Beyond simulation, we use measured DICHASUS channels \citep{euchner2021dichasus} to evaluate adaptation with limited measured data, where RFWeave performs effectively with as few as 100 samples. We also evaluate RFWeave on a downstream beam management task in a simulated factory \citep{bian2026gsbf}, where it enables beam decisions that achieve 34.5\% higher rates than location-based beam steering. Overall, our key contributions are summarized as follows:

\begin{itemize}
    \item We propose RFWeave, a unified feedforward model for wireless propagation prediction across scenes, transceiver positions, and transceiver configurations.

    \item We develop a flexible antenna representation to capture antenna element geometry, together with a conditioning mechanism that enables a shared Transformer backbone to accommodate diverse transceiver configurations.
   
    \item Through extensive experiments, we demonstrate that RFWeave achieves state-of-the-art performance across diverse generalization settings and validate its adaptability to real-world channels and downstream wireless tasks.
\end{itemize}

\begin{figure*}[t]
\centering
\includegraphics[width=0.95\textwidth]{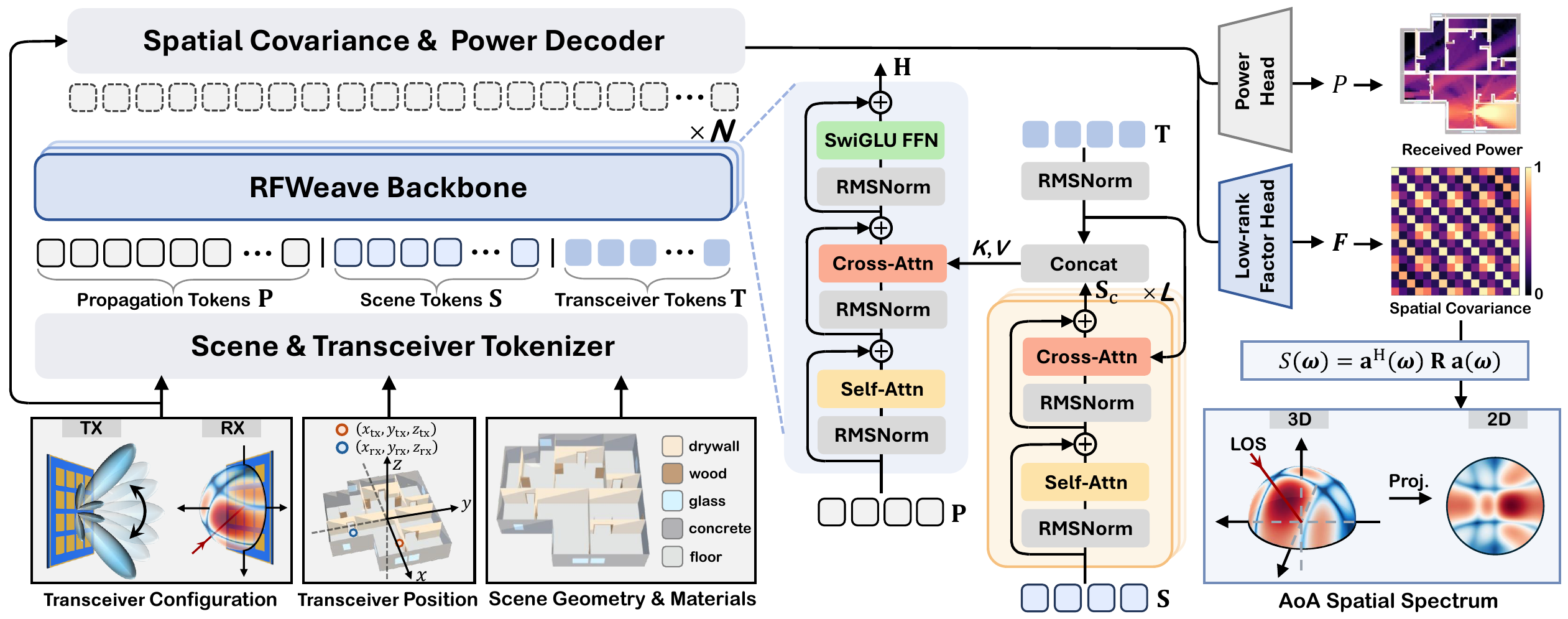}
\caption{\textbf{RFWeave model overview.} RFWeave organizes scene and transceiver information into structured tokens and refines element-indexed propagation tokens with a scene-conditioned backbone. The configuration-conditioned decoder predicts received power and a complex covariance factor, from which the spatial covariance and the angle-of-arrival (AoA) spatial spectrum are derived. For visualization, the spatial spectrum is sampled on a $360^\circ\times90^\circ$ azimuth--elevation grid and projected onto the horizontal plane.}\label{fig:rfweave_overview}
\end{figure*}

\section{Related Work}

\paragraph{Learned wireless propagation modeling.}
Learned propagation modeling has been studied under both scene-specific and cross-scene settings. Inspired by neural radiance fields \citep{mildenhall2020nerf} and 3D Gaussian Splatting \citep{kerbl2023gaussians}, scene-specific approaches reconstruct wireless radiation fields from scene geometry, sparse measurements, or both, enabling propagation prediction at unobserved positions \citep{zhao2023nerf2,lu2024newrf,wen2025wrfgs,yang2025gsrf,zhang2026rf3dgs,shen2026gaigs,umer2025ngrf}. Other approaches learn electromagnetic properties of the environment and integrate them with differentiable ray tracing \citep{hoydis2023sionnart}, but still require detailed 3D scene models and incur the computational cost of ray tracing \citep{chen2025rfscape,ai2026wedt}. These formulations are typically built or adapted for a target environment. Cross-scene methods instead learn transferable propagation models from multiple environments and predict wireless characteristics in unseen scenes \citep{levie2021radiounet,wang2025radiodiff,wang2026radiodiffk2,shen2026gaigs,hehn2025wigatr,yang2026graf}. In particular, Wi-GATr builds on the Geometric Algebra Transformer \citep{brehmer2023gatr} to capture geometric relationships between the scene and transceivers. However, these methods typically assume a common transceiver configuration across scenes, so the effects of antenna configuration are learned implicitly under a fixed setup. We consider the more general setting where scene, transceiver positions, and configurations vary jointly.

\para{Transceiver-aware wireless learning.}
Transceiver variation has mainly been studied in downstream communication tasks. Prior approaches incorporate receiver orientation or multi-panel antenna structure for beam prediction and selection \citep{rezaie2022orientation_beam,rezaie2022multipanel}. More recent work studies beam selection under heterogeneous antenna configurations, including variations in array size, orientation, codebooks, element patterns, and polarization \citep{rezaie2022deviceagnostic, kilinc2026heterogeneity}. Environment-aware approaches also use 3D scene representations to support beam prediction and beamforming \citep{klautau2019lidar,xu2020scene_beam,bian2026gsbf}. These methods account for transceiver variation, but primarily target downstream tasks rather than wireless propagation modeling.

\section{Method}\label{sec:method}

\subsection{Problem Formulation}

We consider a learned forward model of wireless propagation in which the receive-side response depends jointly on the scene $\mathcal{S}$, transceiver positions $\mathcal{P}$, and transceiver configuration $\mathcal{C}$. The transmit array contains $N_{\mathrm{t}}$ antenna elements, while the receive array contains $N_{\mathrm{r}}$ antenna elements.

We characterize the receive-side response through its spatial covariance. Under wide-sense stationary channel models, second-order spatial statistics capture correlations across array elements and relate closely to the angular distribution of received energy \citep{yang2023structured}. Spatial covariance also avoids requiring a consistent absolute phase reference for instantaneous channel realizations. We normalize the covariance to separate its spatial structure from the overall signal scale, and model the latter through received power. With learnable parameters $\theta$, RFWeave learns the mapping
\begin{equation}
    f_{\theta}(\mathcal{S},\mathcal{P},\mathcal{C})
    \mapsto
    \left(\mathbf R, P\right).
    \label{eq:problem}
\end{equation}
Here, ${\mathbf R}\in\mathbb{C}^{N_{\mathrm{r}}\times N_{\mathrm{r}}}$ and $P$ denote the normalized receive-side spatial covariance and received power, respectively. The spatial spectrum is obtained from ${\mathbf R}$ through steering vector evaluation, as described in \cref{sec:decoder}. Our goal is a unified model that generalizes across scenes, transceiver positions, and transceiver configurations, including joint variations across these factors.

\subsection{Scene and Transceiver Tokenization}\label{sec:tokenization}
Geometry relative to the scene is expressed in a common global frame, whereas array-specific quantities such as element positions and beam directions are defined in local array frames. We use a global scene frame $G$, and let $T$ and $R$ denote local transmitter and receiver frames anchored at the corresponding array centers. Accordingly, ${}^{G}\mathbf p$, ${}^{T}\mathbf p$, and ${}^{R}\mathbf p$ denote positions expressed in $G$, $T$, and $R$.

\para{Scene tokenizer.}
We represent the scene as a material-annotated point cloud $\mathcal{S}=\{(\mathbf{x}_k,\mathbf{n}_k,\mathbf{u}_k)\}_{k=1}^{N_{\mathrm{p}}}$ with $N_{\mathrm{p}}$ points, each point described by its position $\mathbf{x}_k \in {\mathbb R}^3$ and surface normal $\mathbf{n}_k\in\mathbb{R}^3$ in the global frame. The electromagnetic material properties $\mathbf{u}_k={(\epsilon_{\mathrm{r}},\log_{10}\sigma)}$ consist of the relative permittivity $\epsilon_{\mathrm{r}}$ and the conductivity $\sigma$. We employ a PointNet++-style set abstraction hierarchy \citep{qi2017pointnet,qi2017pointnetpp}, a widely used approach for hierarchical point cloud representation that has also been adopted in wireless learning tasks \citep{wang2026pointcloudckm,li2026pointcloudmaterial}. At each set abstraction stage, representative points are selected using farthest point sampling, and local features are aggregated through attention pooling \citep{zhao2021pointtransformer,lee2019settransformer}:
\begin{equation}
    \mathbf{f}_i^{(\ell)}
    = \operatorname{AttnPool}_{j\in\mathcal{N}_i}
    \left\{\operatorname{MLP}_\ell\left(
    [\mathbf{x}_j-\mathbf{x}_i;\mathbf{f}_j^{(\ell-1)}]
    \right)\right\},
    \label{eq:set_abstraction}
\end{equation}
where $\mathcal{N}_i$ is the neighborhood of the $i$-th sampled point, $\operatorname{MLP}_\ell(\cdot)$ is a learned feature expansion at stage $\ell$, and $\mathbf{f}_i^{(\ell)}$ is its feature, with $\mathbf{f}_k^{(0)}$ initialized from $\mathbf{x}_k$, $\mathbf{n}_k$, and $\mathbf{u}_k$.

Three abstraction stages progressively downsample the point set by factors of 4, 4, and 2, respectively. The resulting $N_{\mathrm{s}}$ local features are projected into the common token dimension $d$ as $\mathbf{s}_i=\operatorname{Proj}(\mathbf{f}_i^{(3)})$. Stacking these tokens yields $\mathbf{S}\in\mathbb{R}^{N_{\mathrm{s}}\times d}$, which we refer to as the \textit{scene tokens}.

\para{Transceiver tokenizer.}
To accommodate varying array sizes and element arrangements, we represent each antenna element as an individual token using its position in the corresponding local array frame, normalized by the carrier wavelength $\lambda$. Specifically, ${}^{T}\mathbf{p}_i/\lambda$ and ${}^{R}\mathbf{p}_j/\lambda$ denote the normalized positions of the $i$-th TX and $j$-th RX elements, respectively. Following the TX--RX link tokenization in Wi-GATr \citep{hehn2025wigatr}, we additionally introduce three tokens representing the TX and RX array centers in the global frame and their center-to-center displacement. To distinguish different token roles, we use six-dimensional one-hot type vectors, with five types for TX elements, RX elements, TX centers, RX centers, and the TX--RX link, and one for the scene tokens defined above. Each transceiver-related token is obtained by projecting its geometric features and type encoding into the common token dimension $d$:
\begin{equation}
    \mathbf{t}_k=\operatorname{Proj}(
    [\mathbf{g}_k;\text{one-hot}_k]),
    \quad k=1,\ldots,N_{\mathrm{t}}+N_{\mathrm{r}}+3,
    \label{eq:transceiver_token_projection}
\end{equation}
where $\mathbf{g}_k\in\mathbb{R}^3$ denotes the corresponding geometric input (e.g., $\mathbf{g}_i={}^{T}\mathbf{p}_i/\lambda$ for the $i$-th transmitter element). Stacking these tokens forms a variable-length sequence $\mathbf{T}\in\mathbb{R}^{(N_{\mathrm{t}}+N_{\mathrm{r}}+3)\times d}$, naturally accommodating different array sizes. We refer to them as the \textit{transceiver tokens}.

\subsection{Wireless Propagation Modeling}\label{sec:pairwise}

\paragraph{Pairwise geometry encoding.}
While the transceiver tokens capture geometry at the element, array, and link levels, they do not represent the pairwise geometry between individual TX and RX elements. For each TX element $i$ and RX element $j$, we compute their global positions ${}^{G}\mathbf{p}_i^{\mathrm{t}}$ and ${}^{G}\mathbf{p}_j^{\mathrm{r}}$ under fixed reference orientations, and define their relative displacement as $\boldsymbol{\delta}_{ij}={}^{G}\mathbf{p}_j^{\mathrm{r}}-{}^{G}\mathbf{p}_i^{\mathrm{t}}$. A learned projection then encodes their wavelength-normalized local positions $({}^{T}\mathbf{p}_i/\lambda,{}^{R}\mathbf{p}_j/\lambda)$, global positions $({}^{G}\mathbf{p}_i^{\mathrm{t}},{}^{G}\mathbf{p}_j^{\mathrm{r}})$, relative displacement $\boldsymbol{\delta}_{ij}/\lambda$, distance $\|\boldsymbol{\delta}_{ij}\|/\lambda$, and unit direction $\boldsymbol{\delta}_{ij}/\|\boldsymbol{\delta}_{ij}\|$ into a pairwise feature $\mathbf{e}_{ij}\in\mathbb{R}^d$. Collecting these features yields a tensor $\mathbf{E}=[\mathbf{e}_{ij}]_{i=1,j=1}^{N_{\mathrm{t}},N_{\mathrm{r}}}$.

However, keeping all $N_{\mathrm{t}}N_{\mathrm{r}}$ pairwise features throughout the backbone incurs substantial memory and computational costs for large arrays. We aggregate $\mathbf E$ along both the TX and RX dimensions. For each TX element $i$, a TX-indexed query $\mathbf{q}_i^{\mathrm{t}} \in \mathbb{R}^{d}$ attends to $\mathbf E_{i,:} \in \mathbb{R}^{N_{\mathrm{r}} \times d}$ as keys and values to summarize its pairwise relations with all RX elements, while the RX side is constructed symmetrically using RX-indexed query $\mathbf{q}_j^{\mathrm{r}} \in \mathbb{R}^{d}$ from the columns $\mathbf E_{:,j} \in \mathbb{R}^{N_{\mathrm{t}} \times d}$:  
\begin{equation}
    \begin{aligned}
        \mathbf{z}_i^{\mathrm{t}}
        &=\operatorname{Attn}\left(
        \mathbf{W}_\mathrm{Q}^{\mathrm{t}}\mathbf{q}_i^{\mathrm{t}},
        \mathbf{W}_\mathrm{K}^{\mathrm{t}}\mathbf{E}_{i,:}^{\mathsf T},
        \mathbf{W}_\mathrm{V}^{\mathrm{t}}\mathbf{E}_{i,:}^{\mathsf T}
        \right),\\
        \mathbf{z}_j^{\mathrm{r}}
        &=\operatorname{Attn}\left(
        \mathbf{W}_\mathrm{Q}^{\mathrm{r}}\mathbf{q}_j^{\mathrm{r}},
        \mathbf{W}_\mathrm{K}^{\mathrm{r}}\mathbf{E}_{:,j}^{\mathsf T},
        \mathbf{W}_\mathrm{V}^{\mathrm{r}}\mathbf{E}_{:,j}^{\mathsf T}
        \right).
    \end{aligned}
    \label{eq:pairwise_aggregation}
\end{equation}
where $\mathbf{z}_i^{\mathrm{t}}$ and $\mathbf{z}_j^{\mathrm{r}}$ denote the aggregated features, and $\mathbf{W}_\mathrm{Q}$, $\mathbf{W}_\mathrm{K}$, and $\mathbf{W}_\mathrm{V}$ denote the learned projection matrices. These bidirectional aggregated features form $N_{\mathrm{t}}+N_{\mathrm{r}}$ element-indexed \textit{propagation tokens} $\mathbf{P}\in\mathbb{R}^{(N_{\mathrm{t}}+N_{\mathrm{r}})\times d}$. 

\para{Scene conditioning.}
As shown in \cref{fig:rfweave_overview}, to incorporate scene context, we first refine the scene tokens $\mathbf{S}$ through self-attention, after which they attend to the transceiver tokens $\mathbf T$ through cross-attention, yielding a scene representation $\mathbf{S}_{\mathrm{c}}$ conditioned on the current link and array geometry. The propagation tokens $\mathbf{P}$ are then fed into the Transformer backbone. Each Transformer block first models interactions among propagation tokens through self-attention, followed by cross-attention to the concatenated $\mathbf T$ and $\mathbf S_{\mathrm c}$:
\begin{equation}
    \mathbf{H}^{(\ell+1)}
    =\operatorname{CrossAttn}\bigl(
    \operatorname{SelfAttn}(\mathbf{H}^{(\ell)}),
    [\mathbf{T};\mathbf{S}_{\mathrm{c}}]\bigr),
    \label{eq:backbone}
\end{equation}
where $\mathbf{H}^{(0)}=\mathbf{P}$ and $\ell$ indexes the Transformer blocks. The backbone outputs $\mathbf{H}\in\mathbb{R}^{(N_{\mathrm{t}}+N_{\mathrm{r}})\times d}$, consisting of the \textit{TX-indexed tokens} $\{\mathbf{h}_i^{\mathrm{t}}\}_{i=1}^{N_{\mathrm{t}}}$ and \textit{RX-indexed tokens} $\{\mathbf{h}_j^{\mathrm{r}}\}_{j=1}^{N_{\mathrm{r}}}$. Since the queried array orientations and transmit beam are excluded from the backbone, $\mathbf{H}$ can be reused across multiple configuration queries for the same scene, transceiver positions, and array geometry.

\begin{figure*}[t]
\centering
\includegraphics[width=\textwidth]{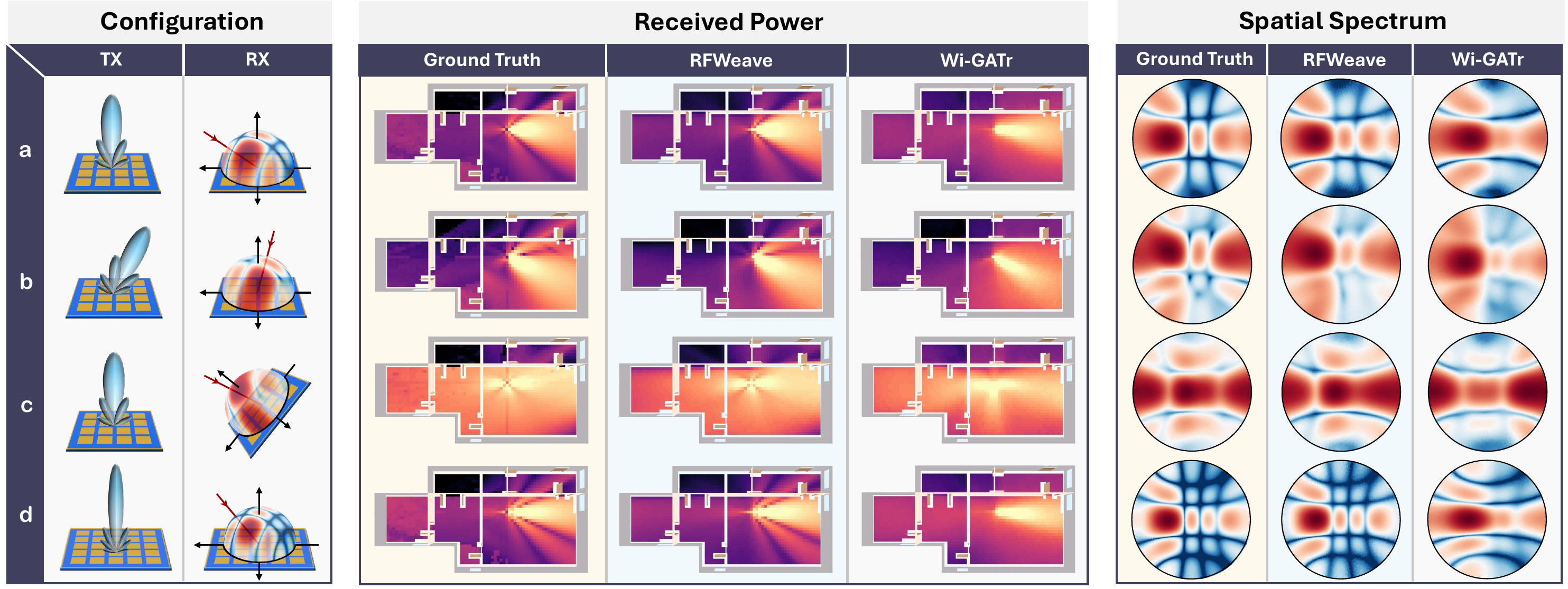}
\caption{\textbf{Comparison on an unseen scene.} We compare received power and spatial spectrum from ground truth, RFWeave, and Wi-GATr \citep{hehn2025wigatr} under different transceiver configurations. \textbf{(a)} uses transmit beam index 0, receiver orientation $180^\circ$, and a $4\times4$ UPA. Panels \textbf{(b)}--\textbf{(d)} vary one configuration factor from this reference: \textbf{(b)} changes the transmit beam index to 5, \textbf{(c)} changes the receiver orientation to $45^\circ$, and \textbf{(d)} increases the array size to $5\times5$. Further analysis is provided in \cref{sec:experimental-results}.}\label{fig:qualitative}
\end{figure*}

\subsection{Spatial Covariance and Power Prediction}\label{sec:decoder}

\paragraph{Configuration conditioning.}
At the prediction stage, we condition the backbone tokens on the queried array orientations and transmit beam. Let ${}^{G}\mathbf{R}_T$ and ${}^{G}\mathbf{R}_R$ denote the queried TX and RX array rotations,  with corresponding forward and up directions denoted by$({}^{G}\mathbf{f}_T,{}^{G}\mathbf{u}_T)$ for the TX array and $({}^{G}\mathbf{f}_R,{}^{G}\mathbf{u}_R)$ for the RX array. The RX-indexed tokens are conditioned as
\begin{equation}
    (\mathbf{h}_j^{\mathrm{r}})^\prime
    =\mathbf{h}_j^{\mathrm{r}}+
    \operatorname{Proj}\Bigl(\Bigl[
    \frac{{}^{R}\mathbf{p}_j}{\lambda};
    \frac{{}^{G}\mathbf{R}_R{}^{R}\mathbf{p}_j}{\lambda};
    {}^{G}\mathbf{f}_R;{}^{G}\mathbf{u}_R
    \Bigr]\Bigr).
    \label{eq:configuration_conditioning}
\end{equation}
The TX-indexed tokens are conditioned analogously to obtain $(\mathbf{h}_i^{\mathrm{t}})^\prime$, with additional inputs $\operatorname{Re}(w_i)$, $\operatorname{Im}(w_i)$, and ${}^{G}\mathbf{R}_T{}^{T}\mathbf{b}$, where $w_i$ is the complex beamforming weight applied to TX element $i$, and ${}^{T}\mathbf{b}$ is the transmit beam direction in the local TX frame. The conditioned TX-indexed tokens $(\mathbf{h}_i^{\mathrm{t}})^\prime$ are first refined through self-attention. Then the conditioned RX-indexed tokens $(\mathbf{h}_j^{\mathrm{r}})^\prime$ attend to the refined TX-indexed tokens through cross-attention, followed by self-attention, and finally yield the RX tokens $(\mathbf{h}_j^{\mathrm{r}})^{\prime\prime}$, which are used to predict the receive-side response.

\para{Covariance and power decoding.}
A shared row-wise factor head maps each final RX token $(\mathbf{h}_j^{\mathrm{r}})^{\prime\prime}$ to one row of a complex factor $\mathbf F\in\mathbb C^{N_{\mathrm{r}}\times r}$, where $r$ is a fixed rank budget. Each row is represented by $2r$ real values corresponding to its real and imaginary parts. Directly predicting the spatial covariance requires $O(N_{\mathrm{r}}^2)$ outputs without guaranteeing a Hermitian positive semidefinite covariance. We instead normalize the factor as $\overline{\mathbf F}=\mathbf F/\|\mathbf F\|_{F}$ and form ${\mathbf R}=\overline{\mathbf F}\,\overline{\mathbf F}^{\mathsf H}$. This gives a Hermitian positive semidefinite, unit-trace covariance using only $O(N_{\mathrm{r}}r)$ factor outputs. Given a steering vector $\mathbf a(\boldsymbol\omega)$, the AoA spatial spectrum is evaluated as $S(\boldsymbol\omega)=10\log_{10}\!\left(\mathbf a^{\mathsf H}(\boldsymbol\omega){\mathbf R}\mathbf a(\boldsymbol\omega)\right)$ and peak-normalized by subtracting its maximum over the queried directions. Predicting ${\mathbf R}$ rather than spectrum values on a fixed angular grid also allows the spectrum to be evaluated at any desired set of steering directions. A separate power head mean-pools the final RX tokens to predict the received power $P$.

\begin{table*}[t]
\caption{\textbf{Cross-scene propagation prediction on WiPTR and Wi3R.} The upper block evaluates feedforward prediction, while the lower block allows 100 target-scene samples for adaptation. \appref{app:finetuning} details the fine-tuning protocol and step sensitivity.}
\label{tab:scene-generalization}
\centering
\RFwidetablefont
\RFwidetablespacing
\setlength{\tabcolsep}{0.30pt}
\begin{tabularx}{\linewidth}{@{}l c *{12}{Y}@{}}
\toprule
\multirow[c]{3}{*}{\raisebox{-2.55ex}{Method}} & \multirow[c]{3}{*}{\raisebox{-3.55ex}{\makecell[c]{Per-scene\\Opt.}}} & \multicolumn{6}{c}{WiPTR} & \multicolumn{6}{c}{Wi3R} \\
\cmidrule(lr){3-8}\cmidrule(lr){9-14}
& & \multicolumn{3}{c}{\RFwidegroup{Held-out ID Scenes}} & \multicolumn{3}{c}{\RFwidegroup{OOD Scenes}} & \multicolumn{3}{c}{\RFwidegroup{Held-out ID Scenes}} & \multicolumn{3}{c}{\RFwidegroup{OOD Scenes}} \\
\cmidrule(lr){3-5}\cmidrule(lr){6-8}\cmidrule(lr){9-11}\cmidrule(lr){12-14}
& & \RFwidehSSIM & \RFwidehSpec & \RFwidehPwr & \RFwidehSSIM & \RFwidehSpec & \RFwidehPwr & \RFwidehSSIM & \RFwidehSpec & \RFwidehPwr & \RFwidehSSIM & \RFwidehSpec & \RFwidehPwr \\
\midrule
\rowcolor{RFGrayRow}\multicolumn{14}{@{}l}{\textit{+ Proposed Decoder}}\\
\addlinespace[1pt]
PointNet++~\citep{qi2017pointnetpp} & None & 0.7988 & 3.16 & 4.75 & 0.7907 & 3.26 & 4.51 & 0.7414 & 4.00 & 8.61 & 0.7073 & 4.43 & 10.91 \\
PointNeXt~\citep{qian2022pointnext} & None & 0.7993 & 3.16 & 4.34 & 0.7924 & 3.25 & 4.37 & 0.7421 & 3.96 & 8.60 & 0.7092 & 4.38 & 10.87 \\
Wi-GATr~\citep{hehn2025wigatr} & None & 0.7996 & 3.13 & 3.92 & 0.7963 & 3.14 & 4.02 & 0.7371 & 4.03 & 8.62 & 0.5821 & 5.77 & 11.46 \\
Transformer~\citep{vaswani2017attention} & None & 0.7865 & 3.26 & 4.61 & 0.7803 & 3.36 & 5.14 & 0.7327 & 4.05 & 8.94 & 0.6998 & 4.52 & 11.43 \\
\RFhighlight\RFmethod{RFWeave-B} & None & \RFbest{0.8708} & \RFbest{2.23} & \RFbest{2.75} & \RFbest{0.8597} & \RFbest{2.34} & \RFbest{3.39} & \RFbest{0.7711} & \RFbest{3.63} & \RFbest{8.30} & \RFbest{0.7433} & \RFbest{4.09} & \RFbest{10.13} \\
\midrule
\rowcolor{RFGrayRow}\multicolumn{14}{@{}l}{\textit{+ 100 Samples}}\\
\addlinespace[1pt]
NeRF$^2$~\citep{zhao2023nerf2} & Required & 0.6710 & 3.43 & 1.51 & 0.6408 & 3.71 & 2.62 & 0.7297 & 3.19 & 1.84 & 0.6011 & 4.68 & 4.96 \\
WRF-GS~\citep{wen2025wrfgs} & Required & 0.6397 & 3.83 & 4.52 & 0.6058 & 4.06 & 5.77 & 0.6033 & 5.77 & 4.80 & 0.5846 & 6.14 & 5.52 \\
\RFhighlight\RFmethod{RFWeave-B (FT)} & Required & \RFbest{0.9398} & \RFbest{1.50} & \RFbest{0.96} & \RFbest{0.9211} & \RFbest{1.76} & \RFbest{1.97} & \RFbest{0.9385} & \RFbest{1.41} & \RFbest{1.49} & \RFbest{0.8858} & \RFbest{2.07} & \RFbest{2.78} \\
\bottomrule
\end{tabularx}
\end{table*}
\begin{table*}[t]
\caption{\textbf{Single-factor configuration generalization on unseen WiPTR scenes.} The left and right blocks evaluate seen and unseen configurations, respectively. Full spatial spectrum MAE results are provided in \appref{sec:appendix-single-factor}.}
\label{tab:single-factor}
\centering
\RFwidetablefont
\RFwidetablespacing
\setlength{\tabcolsep}{1.30pt}
\begin{tabularx}{\linewidth}{@{}l*{12}{Y}@{}}
\toprule
\multirow[c]{3}{*}[-1ex]{Method} & \multicolumn{6}{c}{Seen Configurations} & \multicolumn{6}{c}{Unseen Configurations} \\
\cmidrule(lr){2-7}\cmidrule(lr){8-13}
& \multicolumn{2}{c}{Array Size} & \multicolumn{2}{c}{Receiver Orientation} & \multicolumn{2}{c}{Transmit Beam} & \multicolumn{2}{c}{Array Size} & \multicolumn{2}{c}{Receiver Orientation} & \multicolumn{2}{c}{Transmit Beam} \\
\cmidrule(lr){2-3}\cmidrule(lr){4-5}\cmidrule(lr){6-7}\cmidrule(lr){8-9}\cmidrule(lr){10-11}\cmidrule(lr){12-13}
& \RFwidehSSIM & \RFwidehPwr & \RFwidehSSIM & \RFwidehPwr & \RFwidehSSIM & \RFwidehPwr & \RFwidehSSIM & \RFwidehPwr & \RFwidehSSIM & \RFwidehPwr & \RFwidehSSIM & \RFwidehPwr \\
\midrule
\rowcolor{RFGrayRow}\multicolumn{13}{@{}l}{\textit{Unseen scenes}}\\
\addlinespace[1pt]
PointNet++~\citep{qi2017pointnetpp} & 0.7583 & 4.84 & 0.7848 & 4.35 & 0.7521 & 4.36 & 0.6086 & 5.73 & 0.7280 & 4.37 & 0.6893 & 7.61 \\
PointNeXt~\citep{qian2022pointnext} & 0.7563 & 4.52 & 0.7909 & 4.09 & 0.7520 & 4.34 & 0.6052 & 5.72 & 0.7302 & 4.10 & 0.6989 & 7.38 \\
Wi-GATr~\citep{hehn2025wigatr} & 0.7633 & 3.93 & 0.7843 & 3.58 & 0.7526 & 3.60 & 0.6193 & 5.53 & 0.6906 & 3.64 & 0.7074 & 6.88 \\
Transformer~\citep{vaswani2017attention} & 0.7411 & 4.71 & 0.7690 & 4.34 & 0.7297 & 4.48 & 0.6112 & 5.79 & 0.7061 & 4.39 & 0.6886 & 7.93 \\
\RFhighlight\RFmethod{RFWeave-B$^{\dagger}$} & 0.8420 & 2.97 & 0.8665 & 2.64 & 0.8477 & 2.71 & 0.6348 & 5.44 & 0.8159 & 2.55 & 0.8035 & 4.97 \\
\RFhighlight\RFmethod{RFWeave-B$^{\ddagger}$} & \RFbest{0.8456} & \RFbest{2.76} & \RFbest{0.8703} & \RFbest{2.36} & \RFbest{0.8552} & \RFbest{2.33} & \RFbest{0.6517} & \RFbest{5.39} & \RFbest{0.8226} & \RFbest{2.41} & \RFbest{0.8141} & \RFbest{4.54} \\
\bottomrule
\end{tabularx}
\end{table*}

\subsection{Training Objectives}

Although the covariance is parameterized through the factor $\mathbf F$, we supervise the spatial spectrum induced by the resulting covariance rather than the factor itself. Direct factor supervision is ambiguous because $\mathbf F$ and $\mathbf F\mathbf U$ yield the same covariance for any unitary matrix $\mathbf U$. An entry-wise covariance loss is also possible, but it constrains matrix coefficients rather than the directional responses used for evaluation. We instead compute $S$ from ${\mathbf R}$ and supervise the resulting spatial spectrum. This preserves ${\mathbf R}$ as a positive semidefinite, grid-independent output. Because the unit-trace normalization removes the covariance scale, the signal level is supervised separately through received power.

We use Smooth-L1 losses $\mathcal{L}_{\mathrm{spec}}$ and $\mathcal{L}_{\mathrm{power}}$ for the peak-normalized AoA spatial spectrum and received power, respectively. We additionally supervise configuration-dependent variation along the receiver orientation and transmit beam axes. Let $S_{\mathrm{gt}}$ denote the target spectrum and $\sigma_c(\cdot)$ the per-direction standard deviation across configuration axis $c$. For $c\in\{\mathrm{rx},\mathrm{beam}\}$, we define $\mathcal{L}_{\mathrm{var},c}=\operatorname{SmoothL1}(\sigma_c(S),\sigma_c(S_{\mathrm{gt}}))$, averaged over the remaining dimensions. The full training objective is
\begin{equation}
    \mathcal{L}
    =
    \mathcal{L}_{\mathrm{spec}}
    +
    \mathcal{L}_{\mathrm{power}}
    +
    \lambda_{\mathrm{rx}}\mathcal{L}_{\mathrm{var,rx}}
    +
    \lambda_{\mathrm{beam}}\mathcal{L}_{\mathrm{var,beam}}.
    \label{eq:objective}
\end{equation}
Here, $\lambda_{\mathrm{rx}}$ and $\lambda_{\mathrm{beam}}$ weight the two variation terms, respectively. Each variation term encourages the predicted spectrum to match the target dispersion across the corresponding configuration axis.

\begin{table*}[t]
\caption{\textbf{Joint configuration generalization on unseen WiPTR scenes.} We evaluate seen configurations, angular interpolation of receiver orientations and transmit beams, array size extrapolation, and joint generalization across all three configuration factors.}
\label{tab:joint-generalization}
\centering
\RFwidetablefont
\RFtablespacing
\setlength{\tabcolsep}{1.00pt}
\begin{tabularx}{\linewidth}{@{}l*{12}{Y}@{}}
\toprule
\multirow[c]{2}{*}[-0.20ex]{Method} & \multicolumn{3}{c}{Seen Configurations} & \multicolumn{3}{c}{Angular Interpolation} & \multicolumn{3}{c}{Array Size Extrapolation} & \multicolumn{3}{c}{Joint Generalization} \\
\cmidrule(lr){2-4}\cmidrule(lr){5-7}\cmidrule(lr){8-10}\cmidrule(lr){11-13}
& \RFwidehSSIM & \RFwidehSpec & \RFwidehPwr & \RFwidehSSIM & \RFwidehSpec & \RFwidehPwr & \RFwidehSSIM & \RFwidehSpec & \RFwidehPwr & \RFwidehSSIM & \RFwidehSpec & \RFwidehPwr \\
\midrule
\rowcolor{RFGrayRow}\multicolumn{13}{@{}l}{\textit{Unseen scenes}}\\
\addlinespace[1pt]
PointNet++~\citep{qi2017pointnetpp} & 0.6952 & 4.41 & 4.74 & 0.6102 & 5.29 & 6.74 & 0.5244 & 6.84 & 5.33 & 0.4958 & 7.41 & 7.25 \\
PointNeXt~\citep{qian2022pointnext} & 0.6968 & 4.40 & 4.55 & 0.6135 & 5.24 & 6.63 & 0.5267 & 6.87 & 5.27 & 0.4983 & 7.39 & 7.11 \\
Wi-GATr~\citep{hehn2025wigatr} & 0.6866 & 4.48 & 3.53 & 0.6064 & 5.33 & 5.95 & 0.5396 & 6.75 & 4.93 & 0.5092 & 7.25 & 6.81 \\
Transformer~\citep{vaswani2017attention} & 0.6712 & 4.69 & 4.58 & 0.5997 & 5.50 & 6.52 & 0.5250 & 6.98 & 5.48 & 0.4898 & 7.52 & 7.36 \\
\RFhighlight\RFmethod{RFWeave-T} & 0.7865 & 3.35 & 2.33 & 0.7063 & 4.42 & 5.28 & 0.5543 & 6.05 & 4.70 & 0.5205 & 6.77 & 6.45 \\
\RFhighlight\RFmethod{RFWeave-S} & 0.7960 & 3.24 & 2.32 & 0.7071 & 4.39 & 5.16 & 0.5682 & 5.86 & 4.67 & 0.5379 & 6.52 & 6.22 \\
\RFhighlight\RFmethod{RFWeave-B} & \RFbest{0.8057} & \RFbest{3.14} & \RFbest{2.28} & \RFbest{0.7114} & \RFbest{4.18} & \RFbest{4.93} & \RFbest{0.5704} & \RFbest{5.81} & \RFbest{4.46} & \RFbest{0.5438} & \RFbest{6.44} & \RFbest{6.10} \\
\bottomrule
\end{tabularx}
\end{table*}
\begin{figure*}[t]
\centering
\includegraphics[width=\textwidth]{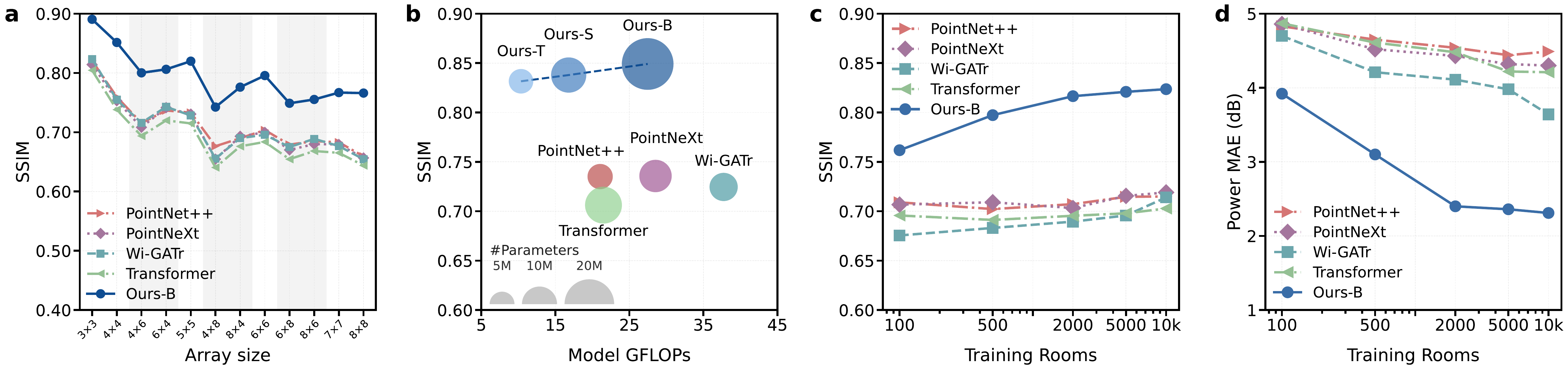}
\caption{\textbf{RFWeave performance analysis.}
\textbf{(a)} Spatial spectrum SSIM across different UPA sizes.
\textbf{(b)} Accuracy--efficiency trade-off across model capacities.
\textbf{(c)} Spatial spectrum SSIM and
\textbf{(d)} received power MAE as the number of training scenes varies.}
\label{fig:scaling_analysis}

\phantomsubcaption\label{fig:array-size-show}
\phantomsubcaption\label{fig:model-capacity-show}
\phantomsubcaption\label{fig:data-efficiency-show}
\phantomsubcaption\label{fig:data-efficiency-power-show}

\end{figure*}
\begin{figure}[t]
\centering
\includegraphics[width=\columnwidth]{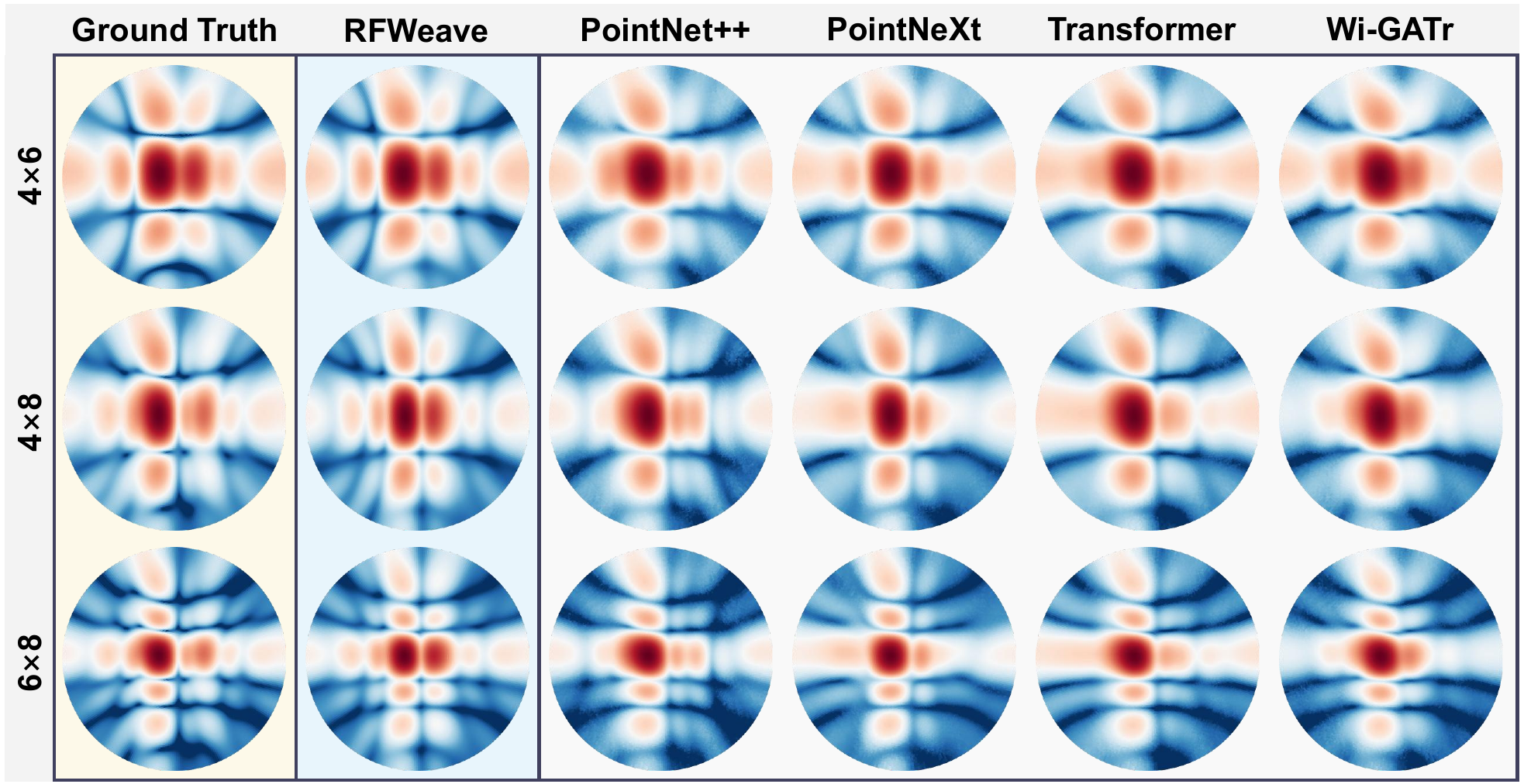}
\caption{Visualization of spatial spectra for rectangular UPAs.}\label{fig:rectangular}
\vspace{-15pt}
\end{figure}

\section{Experiments}

We conduct a series of experiments and ablation studies to answer three questions. (1) How well does RFWeave generalize to unseen scenes under both fixed and varying transceiver configurations? (2) How do key design choices, model capacity, array size, and training data scale affect prediction accuracy and efficiency? (3) Can RFWeave adapt to real-world channel measurements and support downstream beam management? We compare RFWeave against strong baselines in each setting.

\para{Datasets and configurations.}
The main simulated experiments use WiPTR and Wi3R released with Wi-GATr \citep{hehn2025wigatr}. Using far-field synthesis, we construct AoA spatial spectrum and received power targets from their multipath channel records. The resulting data contain 10,000 scenes and over 5 million channel records. Training covers 6 square uniform planar array (UPA) sizes from $3\times3$ to $8\times8$, 8 receiver orientations at $45^\circ$ intervals, and 16 physical transmit beam directions derived from the $4\times4$ UPA DFT codebook. Evaluation further extends to arrays up to $12\times12$, 16 unseen receiver orientations, and 16 unseen transmit beam directions. All test scenes in the main generalization experiments are held out from training. Each antenna element follows the directional pattern in 3GPP TR 38.901 \citep{3gpp38901}. Exact in-distribution (ID) and out-of-distribution (OOD) splits and configuration protocols are provided in \appref{app:generalization}.

\para{Baselines and metrics.}
We compare RFWeave with PointNet++ \citep{qi2017pointnetpp}, PointNeXt \citep{qian2022pointnext}, a vanilla Transformer \citep{vaswani2017attention}, and Wi-GATr \citep{hehn2025wigatr}. Point cloud baselines use the same material-annotated point clouds as RFWeave, while the Transformer and Wi-GATr use material-annotated meshes, following their original implementations. All feedforward baselines use decoders matched to RFWeave in architecture, parameters, and computation, with an additional array size token encoding the number of antenna elements. For adaptation to target scenes, we further compare with NeRF$^2$ \citep{zhao2023nerf2} and WRF-GS \citep{wen2025wrfgs}. We report spatial spectrum SSIM \citep{wang2004ssim} and MAE (dB), and received power MAE (dB), with higher SSIM and lower MAE indicating better performance.

\subsection{Experimental Results and Analysis}\label{sec:experimental-results}

We consider three RFWeave variants: RFWeave-Tiny (T), RFWeave-Small (S), and RFWeave-Base (B). Their parameter counts and detailed accuracy--efficiency trade-offs are reported in \appref{app:scaling}.

\para{Cross-scene generalization.}
We first isolate scene generalization under a fixed transceiver configuration, with the exact configuration and ID/OOD construction provided in \appref{app:cross-scene}. As shown in \cref{tab:scene-generalization}, RFWeave-B consistently outperforms the cross-scene baselines on both WiPTR and Wi3R. All methods degrade on the OOD scenes, particularly on Wi3R, but RFWeave-B remains the strongest feedforward model. For comparison with the scene-specific NeRF$^2$ and WRF-GS baselines, RFWeave-B is fine-tuned using the same 100 target-scene samples and improves on every split. \cref{fig:qualitative} provides a qualitative comparison, where RFWeave better preserves narrow directional power structures and sharp attenuation boundaries while producing a sharper, better-separated spatial spectrum than Wi-GATr.

\para{Single-factor configuration generalization.}
We next vary one configuration factor at a time on unseen WiPTR scenes while keeping other factors fixed. Detailed configuration splits are provided in \appref{app:single-factor}. RFWeave-B$^{\dagger}$ varies only the factor under evaluation during training, whereas RFWeave-B$^{\ddagger}$ is trained jointly across array sizes, receiver orientations, and transmit beams. \cref{tab:single-factor} shows that the two variants remain broadly comparable on seen configurations, while RFWeave-B$^{\ddagger}$ is consistently stronger under unseen configuration changes. Array size extrapolation produces the largest degradation, while RFWeave-B$^{\ddagger}$ remains strongest overall. These results indicate that broader configuration training improves generalization to unseen configurations without sacrificing accuracy on seen ones.

\para{Joint configuration generalization.}
Finally, we consider the more challenging case in which all three configuration factors vary jointly. \cref{tab:joint-generalization} evaluates seen configurations, angular interpolation, array size extrapolation, and joint generalization, with their exact construction detailed in \appref{app:joint-config}. Array size extrapolation causes a larger degradation in spatial spectrum accuracy than angular interpolation, while joint generalization presents the greatest overall challenge. Despite these challenges, RFWeave-B achieves the best results across all four settings. Increasing model capacity from RFWeave-T to RFWeave-B further leads to consistent improvements across these settings.

\para{Effect of array size.}
As shown in \cref{fig:array-size-show}, we evaluate six rectangular UPA sizes ($4\times6$, $6\times4$, $4\times8$, $8\times4$, $6\times8$, and $8\times6$) to examine generalization beyond the square UPAs used during training. RFWeave-B achieves the highest SSIM, while qualitative results are visualized in \cref{fig:rectangular}. We further evaluate square UPAs with side lengths beyond the training range, as detailed in \appref{app:scaling}. All methods exhibit a noticeable performance drop when the side length increases from $N=8$ to $N=9$, crossing the training boundary. Together with \cref{tab:single-factor,tab:joint-generalization}, these results show that RFWeave generalizes well to unseen rectangular arrays with in-range dimensions, while extrapolation to larger arrays remains considerably more challenging.

\para{Model capacity.}
We further examine the trade-off between prediction accuracy and computation. RFWeave-T, RFWeave-S, and RFWeave-B contain 4.89M, 10.08M, and 21.67M parameters, respectively. The model-capacity comparison in \cref{fig:model-capacity-show} reports their accuracy and computational cost. Even RFWeave-T outperforms all the baselines on SSIM at lower computational cost, while RFWeave-S and RFWeave-B further improve accuracy at additional cost. Overall, the three variants provide distinct accuracy--computation trade-offs for different resource budgets.

\para{Data efficiency.}
To examine how performance changes with the amount of training data, we train all methods under the same 20k-step optimization budget using $N_{\mathrm{train}}\in\{100,500,2000,5000,10000\}$ WiPTR rooms, with array size, receiver orientation, and transmit beam jointly sampled within their training ranges. This isolates the effect of training scene diversity from configuration extrapolation. As shown in \cref{fig:data-efficiency-show,fig:data-efficiency-power-show}, RFWeave-B performs best at every evaluated data scale and improves fastest from 100 to 2,000 training rooms. Beyond 2,000 rooms, the gains diminish as performance gradually saturates. The sampling protocol is detailed in \appref{app:scaling}.

\begin{table}[t]
\caption{\textbf{Ablation of the main RFWeave design choices.} $\Delta$ denotes the relative change (\%) from the full model.}
\label{tab:ablation}
\centering
\RFtablefont
\RFtablespacing
\setlength{\tabcolsep}{0.7pt}
\begin{tabularx}{\linewidth}{@{}l*{6}{Y}@{}}
\toprule
\multirow[c]{2}{*}{Variant} & \multicolumn{2}{c}{\RFhSSIM} & \multicolumn{2}{c}{\RFhSpec} & \multicolumn{2}{c}{\RFhPwr}\\
\cmidrule(lr){2-3}\cmidrule(lr){4-5}\cmidrule(lr){6-7}
& Value & $\Delta$ (\%) & Value & $\Delta$ (\%) & Value & $\Delta$ (\%)\\
\midrule
\RFhighlight\RFmethod{RFWeave-B (full)} & \RFbest{0.8057} & -- & \RFbest{3.14} & -- & \RFbest{2.28} & --\\
w/o material properties & 0.7926 & $-1.6$ & 3.37 & $+7.3$ & 2.45 & $+7.5$\\
\rowcolor{RFAltRow}
w/o pairwise geometry & 0.7720 & $-4.2$ & 3.59 & $+14.3$ & 2.74 & $+20.2$\\
\begin{tabular}[c]{@{}l@{}}Configuration conditioning\\$\rightarrow$ orientation/beam tokens\end{tabular} & 0.7533 & $-6.5$ & 3.82 & $+21.7$ & 2.96 & $+29.8$\\
\rowcolor{RFAltRow}
\begin{tabular}[c]{@{}l@{}}Covariance factor\\$\rightarrow$ multipath parameters\end{tabular} & 0.7085 & $-12.1$ & 4.18 & $+33.1$ & 3.32 & $+45.6$\\
\begin{tabular}[c]{@{}l@{}}Spatial spectrum supervision\\$\rightarrow$ covariance supervision\end{tabular} & 0.7794 & $-3.3$ & 3.46 & $+10.2$ & 2.65 & $+16.2$\\
\bottomrule
\end{tabularx}
\end{table}

\subsection{Ablation Studies}
\label{sec:ablations}

\paragraph{Architectural ablations.}
We ablate the main RFWeave design choices in \cref{tab:ablation}. Directly predicting multipath parameters and using them to synthesize the spatial spectrum and received power causes the largest degradation, reducing spatial spectrum SSIM by 12.1\% and increasing received power MAE by 45.6\%. Replacing configuration conditioning with two dedicated tokens, one encoding the receiver orientation and the other encoding the transmit beam direction, reduces SSIM by 6.5\% and increases received power MAE by 29.8\%. Removing pairwise geometry or material properties consistently degrades performance, highlighting the importance of geometric and electromagnetic scene information. Finally, replacing spatial spectrum supervision with spatial covariance supervision degrades all metrics, supporting spectrum-domain supervision.

\para{Configuration variation supervision.}
Each variation term mainly improves its corresponding configuration axis. The balanced setting $(\lambda_{\mathrm{rx}},\lambda_{\mathrm{beam}})=(1,1)$ achieves the best result on five of six metrics and is used in the main experiments. The full sweep is reported in \appref{app:variance-loss}.

\subsection{Additional Evaluations}
\label{sec:additional-evals}
\label{sec:measured}
\label{sec:beam}

\begin{table}[t]
\caption{\textbf{Adaptation to measured DICHASUS channels} with different numbers of training samples.}
\label{tab:dichasus}
\centering
\RFtablefont
\RFtablespacing
\setlength{\tabcolsep}{0.6pt}
\begin{tabularx}{\linewidth}{@{}c*{6}{Y}@{}}
\toprule
\multirow[c]{2}{*}[-0.20ex]{\makecell[c]{Training\\samples}} & \multicolumn{2}{c}{\RFmethod{RFWeave-B (FT)}} & \multicolumn{2}{c}{NeRF$^2$~\citep{zhao2023nerf2}} & \multicolumn{2}{c}{WRF-GS~\citep{wen2025wrfgs}} \\
\cmidrule(lr){2-3}\cmidrule(lr){4-5}\cmidrule(lr){6-7}
& \RFwidehSSIM & \RFwidehPwr & \RFwidehSSIM & \RFwidehPwr & \RFwidehSSIM & \RFwidehPwr \\
\midrule
100  & \RFbest{0.7769} & \RFbest{1.46} & 0.6689 & 2.64 & 0.6428 & 3.03 \\
500  & \RFbest{0.7820} & \RFbest{1.38} & 0.7407 & 1.54 & 0.6433 & 2.76 \\
1000 & \RFbest{0.7836} & \RFbest{1.35} & 0.7443 & 1.47 & 0.6662 & 2.53 \\
5000 & \RFbest{0.7921} & 1.31 & 0.7471 & \RFbest{1.26} & 0.6710 & 2.46 \\
\bottomrule
\end{tabularx}
\end{table}
\begin{table}[t]
\caption{\textbf{Beam management with a $4\times4$ DFT codebook.} Mean / median spectral efficiency (SE) in bps/Hz. Top-$K$ retains the $K$ highest-ranked beam pairs and reports the best.}
\label{tab:beam-management}
\centering
\RFtablefont
\RFtablespacing
\setlength{\tabcolsep}{1.0pt}
\begin{tabularx}{\linewidth}{@{}l*{4}{Y}@{}}
\toprule
Strategy & Top-1 & Top-2 & Top-3\\
\midrule
Exhaustive sweep (GT) & 14.64 / 14.86 & -- & --\\
Location-based steering & 7.24 / 7.31 & -- & --\\
PointNet++ (FT)~\citep{qi2017pointnetpp} & 6.83 / 7.15 & 8.20 / 8.34 & 9.01 / 9.92 \\
PointNeXt (FT)~\citep{qian2022pointnext} & 7.37 / 7.73 & 8.67 / 8.88 & 9.56 / 10.57 \\
\RFhighlight\RFmethod{RFWeave-B (FT)} & 9.74 / 10.30 & 10.33 / 11.26 & 10.78 / 11.61 \\
\bottomrule
\end{tabularx}
\end{table}

\paragraph{Measured channel adaptation.}
We test simulation-to-measurement adaptation on the single hallway DICHASUS dataset \citep{euchner2021dichasus}, collected at 3.438 GHz with a moving transmitter and two fixed receive arrays. We sample a material-annotated point cloud from the provided GLB scene model. For each measurement, we estimate the spatial covariance as $\widehat{\mathbf R}=\frac{1}{K}\sum_{k=1}^{K}\mathbf h_k\mathbf h_k^{\mathsf H}$ by averaging over $K$ subcarriers \citep{salzer2005spatial}, where $\mathbf h_k$ denotes the calibrated CSI vector at the $k$-th subcarrier. We select spatially distributed measurements to ensure broad coverage of the scene, while reserving separate samples for testing. Under the same measurement budgets (\cref{tab:dichasus}), RFWeave achieves the highest spatial spectrum SSIM and performs best in the low-data regime.

\para{Downstream beam management.}
By conditioning on transmit beams, RFWeave can be directly applied to downstream beam management. We evaluate a $4\times4$ UPA with its DFT codebook in a simulated factory \citep{bian2026gsbf}. RFWeave-B (FT) substantially outperforms location-based steering, increasing mean Top-1 SE from 7.24 to 9.74 bps/Hz (\cref{tab:beam-management}). Details are provided in \appref{app:beam-selection}.

\section{Conclusion}
We proposed RFWeave, a learned wireless propagation model for joint generalization across scenes, transceiver positions, and transceiver configurations. Variable-length transceiver tokens preserve array structure, while element-indexed propagation tokens encode pairwise geometry between transmitter and receiver elements. Combined with configuration conditioning, these representations enable spatial covariance and received power prediction across array sizes, receiver orientations, and transmit beams. Across WiPTR and Wi3R, RFWeave achieves state-of-the-art performance. It also adapts to DICHASUS measurements and supports downstream beam management tasks. Together, these results demonstrate the potential of unified feedforward models for wireless propagation prediction.
\section*{LLM Usage Statement}
OpenAI GPT and Alibaba Qwen models were used to assist with language polishing, manuscript drafting and revision, literature search, and code development. All AI-assisted content was reviewed and verified by the authors. The authors take full responsibility for the technical content, experimental results, and conclusions presented in this paper.

{
    \small
    \bibliographystyle{ieeenat_fullname}
    \bibliography{main}
}

\clearpage
\maketitlesupplementary
\appendix
\begingroup
\renewcommand{\RFtablefont}{\scriptsize}
\begin{strip}
\centering
\includegraphics[width=0.98\linewidth]{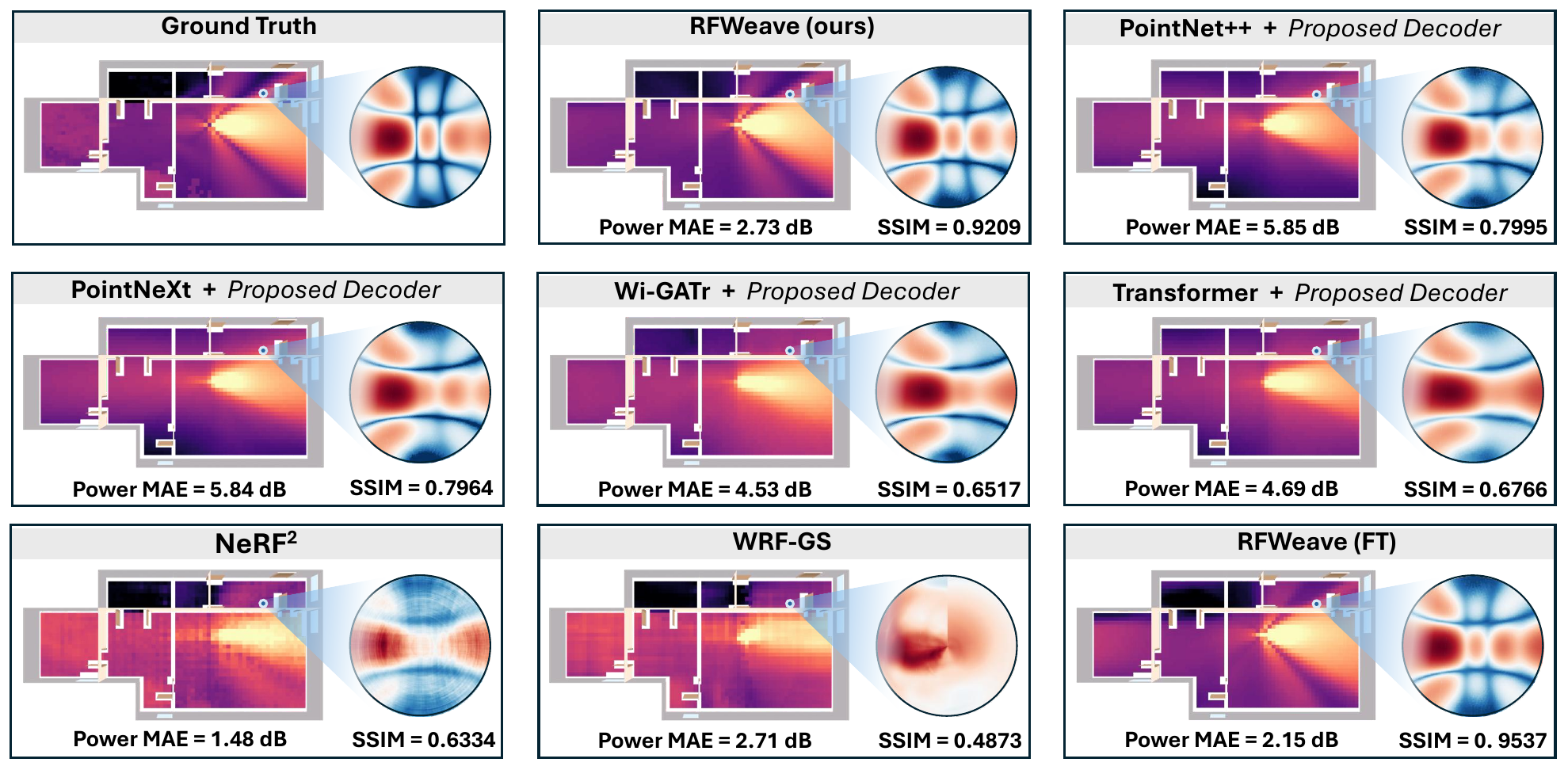}
\captionsetup{hypcap=false}
\captionof{figure}{\textbf{Cross-scene prediction and target-scene adaptation.} Qualitative comparison on an unseen scene. Each method predicts the received power map and the AoA spatial spectrum at the marked receiver location. RFWeave denotes the feed-forward prediction, while RFWeave (FT) is fine-tuned using 100 target-scene samples. NeRF$^2$ and WRF-GS are optimized independently for the target scene using the same samples. RFWeave better preserves the narrow high-power structure and separated spectrum lobes than baselines.}
\captionsetup{hypcap=true}\label{fig:cross-scene-qualitative}
\end{strip}
\begin{figure*}[t]
\centering
\includegraphics[width=0.98\linewidth]{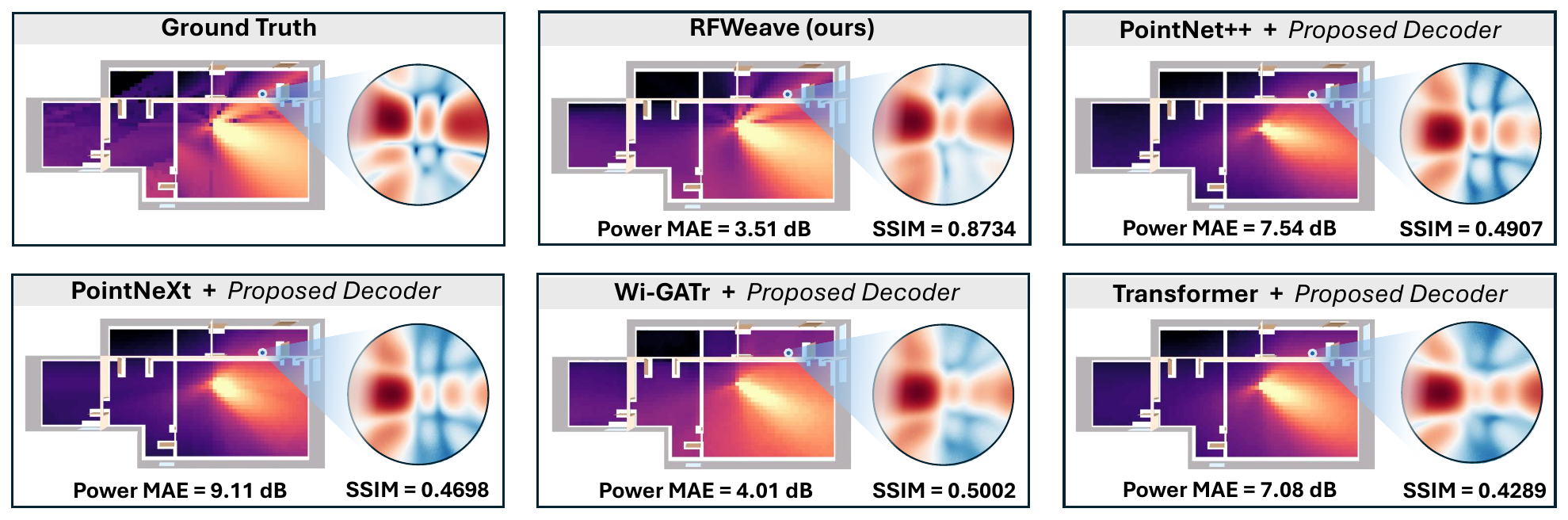}
\caption{\textbf{Transmit beam generalization.} Feedforward predictions on an unseen WiPTR scene as the transmit beam changes from index 0 to 5, while array size and receiver orientation remain fixed.}
\label{fig:beam-qualitative}
\end{figure*}

\begin{figure*}[t]
\centering
\includegraphics[width=0.98\linewidth]{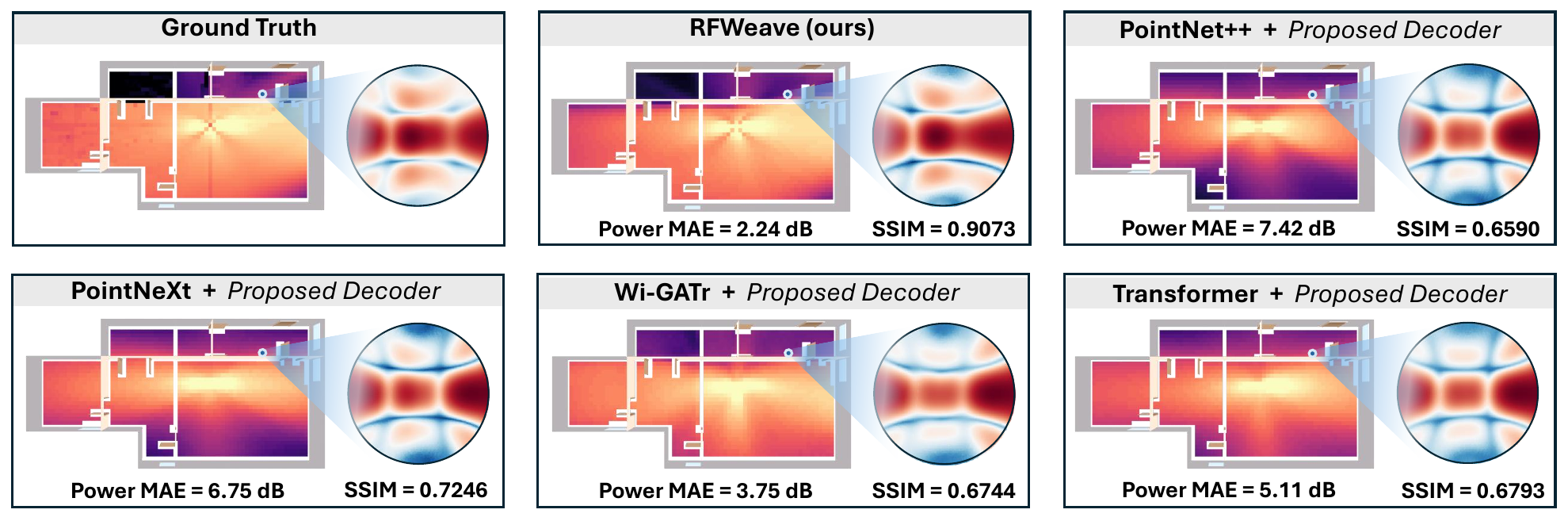}
\caption{\textbf{Receiver orientation generalization.} Feedforward predictions on an unseen WiPTR scene as the receiver orientation changes from the reference orientation to $45^\circ$, while array size and transmit beam remain fixed.}
\label{fig:rx-orientation-qualitative}
\end{figure*}

\section{Implementation Details}
\label{app:implementation}

\subsection{Details of Scene Tokenizer.}
    \phantomsection\label{app:scene-tokenizer}
    We provide additional implementation details for the scene tokenizer introduced in \cref{sec:tokenization}. Each input point is parameterized by its global position $\mathbf{x}_k$, surface normal $\mathbf{n}_k$, and material properties $\mathbf{u}_k=(\epsilon_{\mathrm{r}},\log_{10}\sigma)$, where $\epsilon_{\mathrm{r}}$ and $\sigma$ denote relative permittivity and conductivity, respectively. The two material channels are standardized using statistics from the corresponding training split and encoded by a two-layer MLP. The resulting material embedding is then fused with the geometric features. At each set abstraction stage, the sampled center feature serves as the attention query. Neighboring features are augmented with their relative displacements and Euclidean distances to the center before forming the keys and values. The three stages use neighborhoods of 24, 32, and 64 points, respectively, with a 64-dimensional local representation at each stage. The resulting local representations are combined with a scene one-hot type vector and projected to the common token dimension $d$, forming the material-aware scene tokens $\mathbf S$. Since scene tokenization depends only on the environment, the scene tokens are computed once per scene and reused across all queried transceiver links. \cref{tab:tokenization} summarizes the scene and transceiver tokenization used in RFWeave.

\subsection{Details of Transceiver Tokenizer.}
    \phantomsection\label{app:transceiver-encoding}
    The transceiver tokenizer represents transceiver geometry at the element, array, and link levels. To distinguish token roles, we use six-dimensional one-hot encodings for TX elements, RX elements, TX centers, RX centers, TX--RX links, and scene tokens. A shared learned projection maps the geometric features and type encodings to the common $d$-dimensional token space. For carrier wavelength $\lambda$, the TX and RX element tokens are given by $\mathbf t_i^{\mathrm t}=\operatorname{Proj}([{}^{T}\mathbf p_i/\lambda;\text{one-hot}_{\mathrm{te}}])$ and $\mathbf t_j^{\mathrm r}=\operatorname{Proj}([{}^{R}\mathbf p_j/\lambda;\text{one-hot}_{\mathrm{re}}])$, where the element positions are expressed in their respective local array frames. Following the TX--RX link tokenization in Wi-GATr \citep{hehn2025wigatr}, we denote the TX and RX array center positions in the global frame by ${}^{G}\mathbf c^{\mathrm t}$ and ${}^{G}\mathbf c^{\mathrm r}$, respectively. Their corresponding tokens are $\mathbf t_c^{\mathrm t}=\operatorname{Proj}([{}^{G}\mathbf c^{\mathrm t};\text{one-hot}_{\mathrm{tc}}])$ and $\mathbf t_c^{\mathrm r}=\operatorname{Proj}([{}^{G}\mathbf c^{\mathrm r};\text{one-hot}_{\mathrm{rc}}])$. The link token is defined as $\mathbf t_{\mathrm{link}}=\operatorname{Proj}([{}^{G}\boldsymbol\ell;\text{one-hot}_{\mathrm{link}}])$, where ${}^{G}\boldsymbol\ell={}^{G}\mathbf c^{\mathrm r}-{}^{G}\mathbf c^{\mathrm t}$ denotes the displacement from the TX array center to the RX array center in the global frame. Together, these tokens form the sequence $\mathbf T=[\mathbf t_{1:N_{\mathrm t}}^{\mathrm t};\mathbf t_{1:N_{\mathrm r}}^{\mathrm r};\mathbf t_c^{\mathrm t};\mathbf t_c^{\mathrm r};\mathbf t_{\mathrm{link}}]\in\mathbb R^{(N_{\mathrm t}+N_{\mathrm r}+3)\times d}$. Its length varies naturally with $N_{\mathrm t}$ and $N_{\mathrm r}$, allowing the same tokenizer to accommodate different array sizes.

    \begin{table}[t]
    \caption{\textbf{Scene and transceiver tokenization.} Token roles are distinguished by one-hot encodings.}
    \label{tab:tokenization}
    \centering
    \RFtablefont
    \RFtablespacing
    \setlength{\tabcolsep}{3.5pt}
    \begin{tabular}{@{}llll@{}}
    \toprule
    \parbox[b]{0.16\linewidth}{Data type}
    & \parbox[b]{0.29\linewidth}{Input parameterization}
    & \parbox[b]{0.25\linewidth}{Tokenization}
    & \parbox[b]{0.20\linewidth}{Type encoding}\\
    \midrule

    \parbox[c]{0.16\linewidth}{Scene}
    & \parbox[c]{0.29\linewidth}{$\mathbf x_k$, $\mathbf n_k$, $\mathbf u_k$}
    & \parbox[c]{0.25\linewidth}{\raggedright 1 token per final sampled center}
    & \parbox[c]{0.20\linewidth}{$\text{one-hot}_{\mathrm{scene}}$}\\
    \addlinespace[5pt]

    \parbox[c]{0.16\linewidth}{\raggedright TX/RX elements}
    & \parbox[c]{0.29\linewidth}{${}^{T}\mathbf p_i/\lambda$, ${}^{R}\mathbf p_j/\lambda$}
    & \parbox[c]{0.25\linewidth}{\raggedright 1 token per antenna element}
    & \parbox[c]{0.20\linewidth}{%
        $\text{one-hot}_{\mathrm{te}}$,\newline
        $\text{one-hot}_{\mathrm{re}}$
    }\\
    \addlinespace[5pt]

    \parbox[c]{0.16\linewidth}{Array centers}
    & \parbox[c]{0.29\linewidth}{${}^{G}\mathbf c^{\mathrm t}$, ${}^{G}\mathbf c^{\mathrm r}$}
    & \parbox[c]{0.25\linewidth}{1 token per array}
    & \parbox[c]{0.20\linewidth}{%
        $\text{one-hot}_{\mathrm{tc}}$,\newline
        $\text{one-hot}_{\mathrm{rc}}$
    }\\
    \addlinespace[5pt]

    \parbox[c]{0.16\linewidth}{TX--RX link}
    & \parbox[c]{0.29\linewidth}{${}^{G}\boldsymbol\ell$}
    & \parbox[c]{0.25\linewidth}{1 token per link}
    & \parbox[c]{0.20\linewidth}{$\text{one-hot}_{\mathrm{link}}$}\\
    \bottomrule
    \end{tabular}
\end{table}

    \begin{table*}[t]
    \caption{\textbf{Architecture and parameter counts of the RFWeave variants.} Scene self-attention uses two blocks and the covariance factor rank is $r=16$ for all variants.}
    \label{tab:rfweave-variants}
    \centering
    \small
    \RFtablespacing
    \setlength{\tabcolsep}{5.5pt}
    \begin{tabular*}{0.90\textwidth}{@{\extracolsep{\fill}}lcccccc@{}}
    \toprule
    Model & Hidden dim. & Heads & Prop. stages & Backbone (M) & Decoder (M) & Total (M)\\
    \midrule
    RFWeave-T & 128 & 8 & 6 & 3.62 & 1.27 & 4.89\\
    RFWeave-S & 192 & 8 & 6 & 7.38 & 2.70 & 10.08\\
    RFWeave-B & 256 & 8 & 8 & 16.63 & 5.04 & 21.67\\
    \bottomrule
    \end{tabular*}
\end{table*}

\subsection{Details of Backbone.}
    \paragraph{Bidirectional pairwise aggregation.}
    \phantomsection\label{app:pairwise-aggregation}
    The pairwise tensor $\mathbf E$ is constructed as described in \cref{sec:pairwise}. Before aggregating the pairwise features, the TX center token $\mathbf t_c^{\mathrm t}$ and RX center token $\mathbf t_c^{\mathrm r}$ serve as queries and attend to their respective element tokens $\mathbf t_{1:N_{\mathrm t}}^{\mathrm t}$ and $\mathbf t_{1:N_{\mathrm r}}^{\mathrm r}$, thereby summarizing the overall structure of each array. These summaries are then used to construct element-indexed queries. Specifically, the TX-indexed query $\mathbf q_i^{\mathrm t}$ combines the corresponding TX element token $\mathbf t_i^{\mathrm t}$, the link token $\mathbf t_{\mathrm{link}}$, and the RX array summary, while the RX-indexed query $\mathbf q_j^{\mathrm r}$ is constructed symmetrically from $\mathbf t_j^{\mathrm r}$, $\mathbf t_{\mathrm{link}}$, and the TX array summary. As defined in \cref{eq:pairwise_aggregation}, these queries attend to the corresponding rows and columns of $\mathbf E$, yielding the aggregated features $\mathbf z_i^{\mathrm t}$ and $\mathbf z_j^{\mathrm r}$. The resulting propagation tokens are concatenated as $\mathbf P=[\mathbf z_{1:N_{\mathrm t}}^{\mathrm t};\mathbf z_{1:N_{\mathrm r}}^{\mathrm r}]$. Each pairwise feature contributes through both a TX-indexed row and an RX-indexed column. For square $n\times n$ arrays at both ends, $\mathbf E$ contains $n^4$ pairwise features, whereas the backbone processes only $2n^2$ propagation tokens.

    \para{Architecture and model variants.}
    \phantomsection\label{app:backbone-details}
    The backbone uses two self-attention blocks for scene refinement. Propagation tokens are refined through a sequence of stages. Each stage applies self-attention to the propagation tokens, followed by cross-attention to the transceiver tokens and conditioned scene tokens. The scene conditioning and propagation refinement blocks use pre-normalization with RMSNorm \citep{zhang2019rmsnorm} and QK normalization \citep{henry2020qknorm}. Attention is implemented using FlashAttention \citep{dao2022flashattention}. All variants use multi-head attention with eight heads. Each block further includes a SwiGLU feedforward layer \citep{shazeer2020glu} with an equivalent FFN expansion ratio of 4. RFWeave-T, RFWeave-S, and RFWeave-B follow the same tokenizer, backbone, configuration conditioning, and prediction designs. Model capacity is varied through the hidden dimension and the number of propagation refinement stages, while the decoder width is scaled accordingly. \cref{tab:rfweave-variants} summarizes the model settings for the three RFWeave variants.

\section{Experimental Details}
\label{app:generalization}

\subsection{Datasets and Evaluation Protocols.}

    \paragraph{Dataset construction.}
    The main simulated experiments use the path-level channel records released with WiPTR \citep{hehn2025wigatr} and Wi3R \citep{orekondy2023winert}, both generated at 3.5~GHz. We apply deterministic far-field synthesis to each record to obtain the normalized receive-side spatial covariance, AoA spatial spectrum, and received power for a given transceiver configuration. This allows propagation targets for different array sizes, receiver orientations, and transmit beams to be generated from the same record without rerunning the ray tracer. All square UPAs use half-wavelength element spacing. The 16 training transmit beams are defined by the physical directions of a $4\times4$ UPA DFT codebook, while the beamforming vectors for other array sizes are adjusted to preserve these directions. \cref{tab:configuration-splits} summarizes the configuration protocols used for training and evaluation.
    
    \begin{table}[t]
    \caption{\textbf{Training and evaluation configurations for generalization experiments.} \textit{Seen} denotes values observed during training, \textit{unseen} denotes held-out values, and \textit{fixed} denotes reference values held constant in single-factor experiments.}
    \label{tab:configuration-splits}
    \centering
    \RFtablefont
    \RFtablespacing
    \setlength{\tabcolsep}{2.6pt}
    \begin{tabular}{@{}llll@{}}
    \toprule
    \parbox[c]{0.29\linewidth}{Regime}
    & \parbox[c]{0.23\linewidth}{Array size}
    & \parbox[c]{0.21\linewidth}{Receiver\newline orientation}
    & \parbox[c]{0.18\linewidth}{Transmit beam}\\
    \midrule

    \parbox[c]{0.29\linewidth}{Training}
    & \parbox[c]{0.23\linewidth}{\raggedright 6 seen sizes ($3\times3$--$8\times8$)}
    & \parbox[c]{0.21\linewidth}{\raggedright 8 seen orientations}
    & \parbox[c]{0.18\linewidth}{\raggedright 16 seen directions}\\
    \addlinespace[5pt]

    \parbox[c]{0.29\linewidth}{\raggedright Single-factor: array size}
    & \parbox[c]{0.23\linewidth}{\raggedright 4 unseen sizes ($9\times9$--$12\times12$)}
    & \parbox[c]{0.21\linewidth}{fixed}
    & \parbox[c]{0.18\linewidth}{fixed}\\
    \addlinespace[5pt]

    \parbox[c]{0.29\linewidth}{\raggedright Single-factor: receiver orientation}
    & \parbox[c]{0.23\linewidth}{fixed}
    & \parbox[c]{0.21\linewidth}{\raggedright 16 unseen orientations}
    & \parbox[c]{0.18\linewidth}{fixed}\\
    \addlinespace[5pt]

    \parbox[c]{0.29\linewidth}{\raggedright Single-factor: transmit beam}
    & \parbox[c]{0.23\linewidth}{fixed}
    & \parbox[c]{0.21\linewidth}{fixed}
    & \parbox[c]{0.18\linewidth}{\raggedright 16 unseen directions}\\
    \addlinespace[7pt]

    \parbox[c]{0.29\linewidth}{Joint: seen}
    & \parbox[c]{0.23\linewidth}{\raggedright 6 seen sizes ($3\times3$--$8\times8$)}
    & \parbox[c]{0.21\linewidth}{\raggedright 8 seen orientations}
    & \parbox[c]{0.18\linewidth}{\raggedright 16 seen directions}\\
    \addlinespace[5pt]

    \parbox[c]{0.29\linewidth}{\raggedright Joint: angular interpolation}
    & \parbox[c]{0.23\linewidth}{\raggedright 6 seen sizes ($3\times3$--$8\times8$)}
    & \parbox[c]{0.21\linewidth}{\raggedright 16 unseen orientations}
    & \parbox[c]{0.18\linewidth}{\raggedright 16 unseen directions}\\
    \addlinespace[5pt]

    \parbox[c]{0.29\linewidth}{\raggedright Joint: array size extrapolation}
    & \parbox[c]{0.23\linewidth}{\raggedright 4 unseen sizes ($9\times9$--$12\times12$)}
    & \parbox[c]{0.21\linewidth}{\raggedright 8 seen orientations}
    & \parbox[c]{0.18\linewidth}{\raggedright 16 seen directions}\\
    \addlinespace[5pt]

    \parbox[c]{0.29\linewidth}{\raggedright Joint: joint generalization}
    & \parbox[c]{0.23\linewidth}{\raggedright 4 unseen sizes ($9\times9$--$12\times12$)}
    & \parbox[c]{0.21\linewidth}{\raggedright 16 unseen orientations}
    & \parbox[c]{0.18\linewidth}{\raggedright 16 unseen directions}\\
    \bottomrule
    \end{tabular}
\end{table}

    \begin{table*}[t]
\caption{\textbf{Single-factor configuration generalization on unseen WiPTR scenes.} The upper and lower blocks evaluate seen and unseen configurations, respectively.}
\label{tab:single-factor-complete}
\centering
\RFtablefont
\RFtablespacing
\setlength{\tabcolsep}{1.30pt}
\begin{tabularx}{\linewidth}{@{}l*{9}{Y}@{}}
\toprule
\multirow[c]{2}{*}[-0.20ex]{Method} & \multicolumn{3}{c}{Array Size} & \multicolumn{3}{c}{Receiver Orientation} & \multicolumn{3}{c}{Transmit Beam}\\
\cmidrule(lr){2-4}\cmidrule(lr){5-7}\cmidrule(lr){8-10}
& \RFhSSIM & \RFhSpec & \RFhPwr
& \RFhSSIM & \RFhSpec & \RFhPwr
& \RFhSSIM & \RFhSpec & \RFhPwr\\
\midrule
\rowcolor{RFGrayRow}\multicolumn{10}{@{}l}{\textit{Seen configurations, unseen scenes}}\\
\addlinespace[1pt]
PointNet++  & 0.7583 & 3.63 & 4.84 & 0.7848 & 3.35 & 4.35 & 0.7521 & 3.71 & 4.36\\
PointNeXt   & 0.7563 & 3.65 & 4.52 & 0.7909 & 3.28 & 4.09 & 0.7520 & 3.69 & 4.34\\
Wi-GATr     & 0.7633 & 3.56 & 3.93 & 0.7843 & 3.34 & 3.58 & 0.7526 & 3.68 & 3.60\\
Transformer & 0.7411 & 3.84 & 4.71 & 0.7690 & 3.51 & 4.34 & 0.7297 & 3.93 & 4.48\\
\RFhighlight\RFmethod{RFWeave-B$^{\dagger}$} & 0.8420 & 2.58 & 2.97 & 0.8665 & 2.34 & 2.64 & 0.8477 & 2.53 & 2.71\\
\RFhighlight\RFmethod{RFWeave-B$^{\ddagger}$} & \RFbest{0.8456} & \RFbest{2.54} & \RFbest{2.76} & \RFbest{0.8703} & \RFbest{2.18} & \RFbest{2.36} & \RFbest{0.8552} & \RFbest{2.35} & \RFbest{2.33}\\
\midrule
\rowcolor{RFGrayRow}\multicolumn{10}{@{}l}{\textit{Unseen configurations, unseen scenes}}\\
\addlinespace[1pt]
PointNet++  & 0.6086 & 4.92 & 5.73 & 0.7280 & 3.86 & 4.37 & 0.6893 & 4.33 & 7.61\\
PointNeXt   & 0.6052 & 4.95 & 5.72 & 0.7302 & 3.88 & 4.10 & 0.6989 & 4.28 & 7.38\\
Wi-GATr     & 0.6193 & 4.84 & 5.53 & 0.6906 & 4.26 & 3.64 & 0.7074 & 4.11 & 6.88\\
Transformer & 0.6112 & 5.07 & 5.79 & 0.7061 & 4.14 & 4.39 & 0.6886 & 4.47 & 7.93\\
\RFhighlight\RFmethod{RFWeave-B$^{\dagger}$} & 0.6348 & 4.73 & 5.44 & 0.8159 & 2.82 & 2.55 & 0.8035 & 3.16 & 4.97\\
\RFhighlight\RFmethod{RFWeave-B$^{\ddagger}$} & \RFbest{0.6517} & \RFbest{4.62} & \RFbest{5.39} & \RFbest{0.8226} & \RFbest{2.77} & \RFbest{2.41} & \RFbest{0.8141} & \RFbest{2.95} & \RFbest{4.54}\\
\bottomrule
\end{tabularx}
\end{table*}

    \begin{figure*}[t]
\centering
\includegraphics[width=0.98\linewidth]{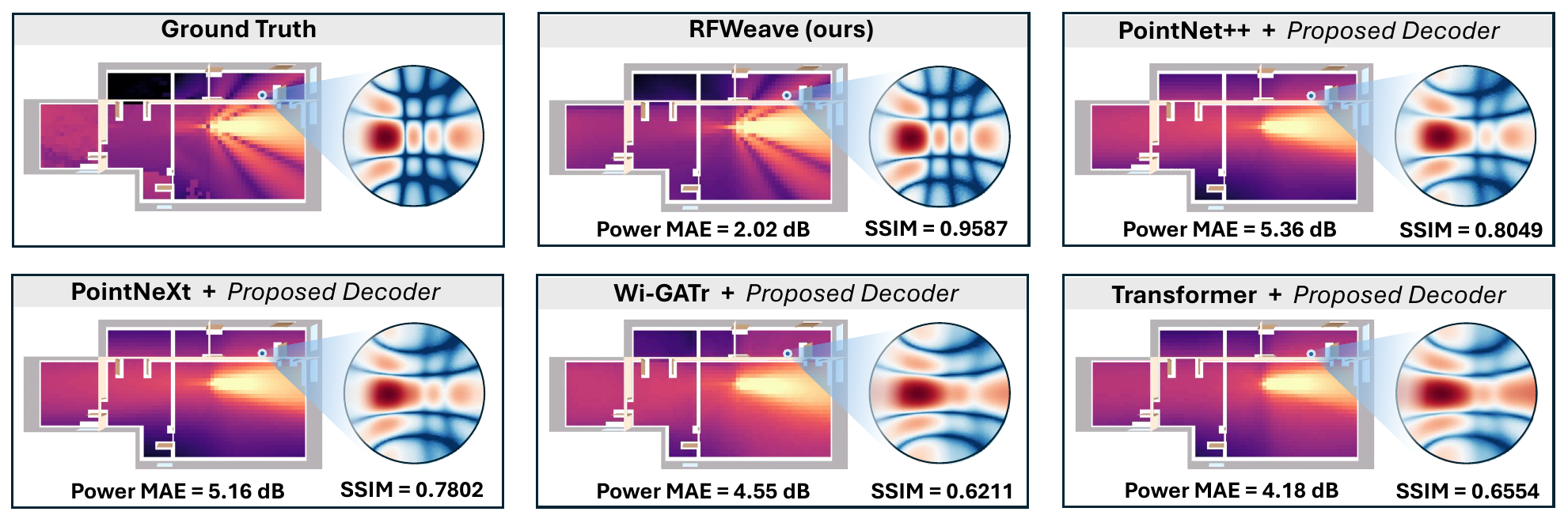}
\caption{\textbf{Array size generalization.} Feedforward predictions on an unseen WiPTR scene as the square UPA changes from $4\times4$ to $5\times5$, while receiver orientation and transmit beam remain fixed.}
\label{fig:array-size-qualitative}
\end{figure*}

    \para{Cross-scene generalization protocol.}
    \phantomsection\label{app:cross-scene}
    We evaluate cross-scene generalization under a fixed transceiver configuration to isolate scene variation. Both TX and RX use a $4\times4$ square UPA. The transmitter uses transmit beam index 0, and the TX and RX orientations are aligned with the global $+x$ and $-x$ axes, respectively. All evaluation scenes are held out from training. For WiPTR, the ID scenes follow the same generation distribution as the training scenes, whereas the OOD scenes alter room connectivity through wall removal. For Wi3R, the ID scenes contain three rooms, whereas the OOD scenes contain four rooms.

    \para{Single-factor configuration generalization protocol.}\label{sec:appendix-single-factor}\label{app:joint-qualitative}
    \phantomsection\label{app:single-factor}
    We evaluate single-factor generalization on unseen WiPTR scenes following the protocols in \cref{tab:configuration-splits}. The 16 unseen transmit beam directions, expressed as (azimuth, elevation) pairs, are
    $(-85^\circ,-35^\circ)$, $(-85^\circ,-10^\circ)$, $(-85^\circ,15^\circ)$, $(-65^\circ,35^\circ)$,
    $(-60^\circ,0^\circ)$, $(-35^\circ,-30^\circ)$, $(-35^\circ,30^\circ)$, $(-30^\circ,-5^\circ)$,
    $(0^\circ,-5^\circ)$, $(5^\circ,-85^\circ)$, $(5^\circ,-30^\circ)$, $(5^\circ,30^\circ)$,
    $(30^\circ,-5^\circ)$, $(60^\circ,-30^\circ)$, $(60^\circ,25^\circ)$, and $(75^\circ,80^\circ)$.
    RFWeave-B$^{\dagger}$ varies only the factor under evaluation during training, while the other two factors remain fixed. In contrast, RFWeave-B$^{\ddagger}$ is trained jointly across array sizes, receiver orientations, and transmit beams. The two variants use the same number of training samples for every complete configuration they share, while RFWeave-B$^{\ddagger}$ additionally covers a broader set of distinct configurations. The full single-factor results, including spatial spectrum MAE, are provided in \cref{tab:single-factor-complete}. \crefrange{fig:cross-scene-qualitative}{fig:array-size-qualitative} provide additional qualitative comparisons for cross-scene prediction and single-factor configuration generalization. The first example additionally includes target-scene adaptation, while the remaining examples show feedforward predictions when one transceiver configuration factor varies at a time.
    
    \para{Joint configuration generalization protocol.}
    \phantomsection\label{app:joint-config}
    All four joint generalization regimes use the same held-out WiPTR scenes, with configuration factors sampled according to \cref{tab:configuration-splits}. Seen configurations vary all three factors within their training sets. Angular interpolation combines seen array sizes with unseen receiver orientations and transmit beam directions. Array size extrapolation combines unseen array sizes ($9\times9$--$12\times12$) with seen orientations and beam directions. Joint generalization evaluates unseen values of all three configuration factors simultaneously. The unseen receiver orientations and transmit beam directions are identical to those used in the single-factor evaluations.

\subsection{Baselines and Optimization.}

\paragraph{Baseline implementations.}
    \phantomsection\label{app:baseline-implementations}
    All reproduced feedforward baselines use the same data splits, optimization budget, and evaluation protocol as RFWeave. PointNet++ \citep{qi2017pointnetpp} and PointNeXt \citep{qian2022pointnext} receive the same material-annotated point clouds as RFWeave, while Wi-GATr \citep{hehn2025wigatr} and the vanilla Transformer \citep{vaswani2017attention} follow the material-annotated mesh representation adopted in the original Wi-GATr. All four baselines use the same covariance and power prediction interface as RFWeave, with matched decoder capacity. To accommodate different array sizes, we append an additional array size token to each baseline to encode the number of antenna elements.

    \para{PointNet++ and PointNeXt.}
    PointNet++ uses a set abstraction encoder with the same point cloud preprocessing and material attributes as RFWeave. PointNeXt follows the official PointNeXt-B configuration. Its input head has width 32, while the four set abstraction stages use stride 4 with output widths of 64, 128, 256, and 512. The corresponding block counts are 2, 3, 2, and 2. Both methods use 3D coordinates for point set geometry operations, while surface normals, relative permittivity, and conductivity are provided as point features.

    \para{Wi-GATr.}
    Our reproduction follows the predictive model setting of \citet{hehn2025wigatr}. It uses 32 geometric-transformer blocks \citep{brehmer2023gatr} with 16 hidden multivector channels, 32 hidden scalar channels, and 8 attention heads. The link, TX, RX, and mesh-face tokens follow the geometric roles defined in the original tokenizer. Scalar channels encode token type and material identity.

    \para{Transformer.}
    The vanilla Transformer uses the same mesh-based input representation as Wi-GATr but removes the geometric-algebra equivariant embedding. It contains 32 Transformer blocks with a hidden dimension of 96 and 8 attention heads. Each block applies multi-head self-attention followed by a feedforward network. The feedforward network uses an MLP ratio of 4 and GELU activations \citep{hendrycks2016gelu}. LayerNorm \citep{ba2016layernorm} is applied within each block.

    \para{Scene-specific baselines.} 
    NeRF$^2$ \citep{zhao2023nerf2} and WRF-GS \citep{wen2025wrfgs} are reconstruction methods optimized independently for each target scene. In the adaptation comparison, both methods use the same 100 target-scene samples provided to RFWeave-B for fine-tuning. RFWeave-B, in contrast, starts from its pretrained checkpoint and adapts to the target scene.
    
    \para{Optimization and adaptation.}
    \phantomsection\label{app:training-details}
    All feedforward models are trained with Adam \citep{kingma2015adam} for 20k steps. The learning rate is selected separately for each model by grid search, with $10^{-3}$ yielding the best validation performance across all reported models. Each training step samples three scenes and 32 transceiver links per scene, giving 96 links in total while allowing links from the same scene to share the scene representation. All reproduced results are obtained on NVIDIA H100 GPUs. For target-scene adaptation, RFWeave-B is initialized from the pretrained feedforward checkpoint and fine-tuned for 500 steps using 100 target-scene samples. NeRF$^2$ and WRF-GS are optimized independently for each target scene for 20k and 60k steps, respectively.
\section{Additional Experiments}
\label{app:additional-experiments}

\subsection{Fine-Tuning Sensitivity}
\label{app:finetuning}

To examine whether the adaptation performance of RFWeave-B depends strongly on the fine-tuning duration, we vary the number of fine-tuning steps while keeping the target-scene sample budget fixed at 100. \cref{tab:finetuning-sensitivity} reports the results with 500, 1,000, and 2,000 fine-tuning steps for the scene-specific comparison in \cref{tab:scene-generalization}. Increasing the number of steps does not lead to consistent improvements across the reported metrics. We therefore use 500 fine-tuning steps in the main experiments.

\begin{table}[t]
\caption{\textbf{Fine-tuning sensitivity to the number of optimization steps.} All settings use the same 100 target-scene samples and the fine-tuning configuration adopted in Table~\ref{tab:scene-generalization}.}
\label{tab:finetuning-sensitivity}
\centering
\RFtablefont
\RFtablespacing
\setlength{\tabcolsep}{2.2pt}
\begin{tabularx}{\linewidth}{@{}c*{6}{Y}@{}}
\toprule
\multirow[c]{2}{*}{Steps} & \multicolumn{3}{c}{WiPTR Held-out ID Scenes} & \multicolumn{3}{c}{Wi3R OOD Scenes} \\
\cmidrule(lr){2-4}\cmidrule(lr){5-7}
& \RFhSSIM & \RFwidehSpec & \RFwidehPwr & \RFhSSIM & \RFwidehSpec & \RFwidehPwr \\
\midrule
\RFhighlight 500  & 0.9398 & \RFbest{1.498} & \RFbest{0.962} & \RFbest{0.8858} & \RFbest{2.069} & 2.782 \\
1000 & 0.9372 & 1.576 & 0.970 & 0.8828 & 2.113 & \RFbest{2.115} \\
2000 & \RFbest{0.9407} & 1.556 & 1.045 & 0.8603 & 2.438 & 3.010 \\
\bottomrule
\end{tabularx}
\end{table}

\begin{table}[t]
\begin{center}
\captionof{table}{Model size and total forward computation at $4\times4$ and $8\times8$ array sizes.}
\label{tab:appendix-complexity}
\RFtablefont
\RFtablespacing
\setlength{\tabcolsep}{2.5pt}
\begin{tabular*}{\linewidth}{@{}l@{\extracolsep{\fill}}ccc@{}}
\toprule
Method & Params. (M) & $4\times4$ GFLOPs & $8\times8$ GFLOPs \\
\midrule
PointNet++  & 5.19  & 21.05 & 67.49 \\
PointNeXt   & 8.52  & 28.53 & 74.97 \\
Wi-GATr     & 6.47  & 37.73 & 84.16 \\
Transformer & 8.22  & 21.50 & 67.93 \\
RFWeave-T   & 4.89  & 10.36 & 28.99 \\
RFWeave-S   & 10.08 & 16.80 & 47.90 \\
RFWeave-B   & 21.67 & 27.47 & 79.80 \\
\bottomrule
\end{tabular*}
\end{center}
\end{table}

\subsection{Scaling and Efficiency}
\label{app:scaling}

For the data efficiency study in \cref{fig:data-efficiency-show}, each method is trained for 20k steps using $N_{\mathrm{train}}\in\{100,500,2000,5000,10000\}$ WiPTR scenes. During training, array size, receiver orientation, and transmit beam are sampled jointly within their respective training ranges. All models are evaluated on the same 2,000 deterministically sampled transceiver links from 100 held-out scenes.

For the model capacity and array size studies, the scene representation is computed once, after which a single transceiver link is decoded for all 16 transmit beam directions and 8 receiver orientations. The spatial spectrum is computed on the same $360\times90$ angular grid used throughout evaluation. Total GFLOPs account for scene and backbone processing, prediction decoding, and spatial spectrum computation. Peak GPU memory is measured with 8 transceiver links sharing a single scene representation. \cref{tab:appendix-complexity} reports the total forward computation for all feedforward models at $4\times4$ and $8\times8$ array sizes.

\begin{figure}[t]
\centering
\includegraphics[width=\linewidth]{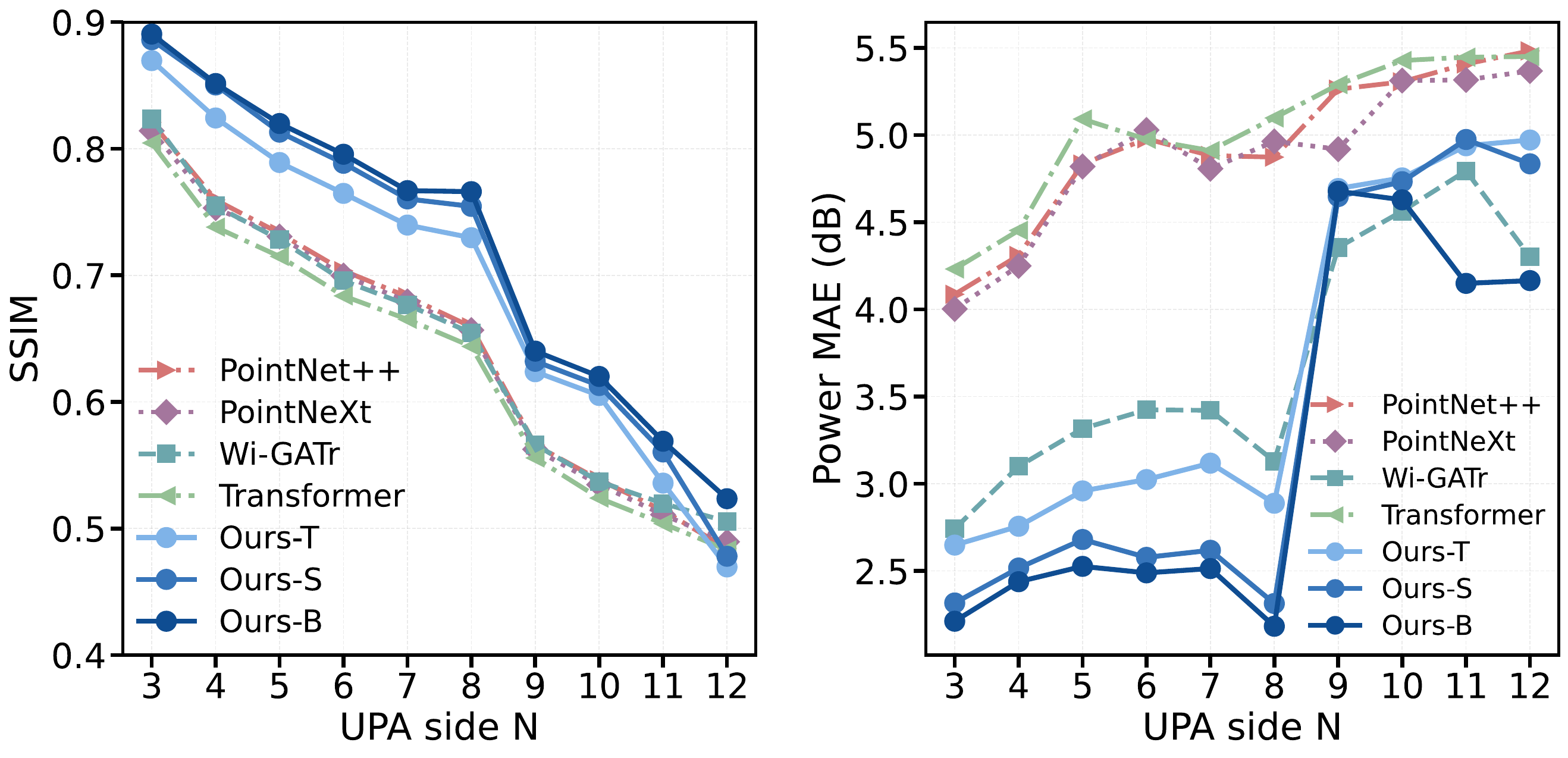}
\caption{\textbf{Spatial Spectrum SSIM and received power MAE across array sizes.} Models are trained on square UPAs with $N=3,\ldots,8$ and additionally evaluated on unseen array sizes $N=9,\ldots,12$.}
\label{fig:array-power-mae}
\end{figure}

\begin{figure}[t]
\centering
\includegraphics[width=0.98\linewidth]{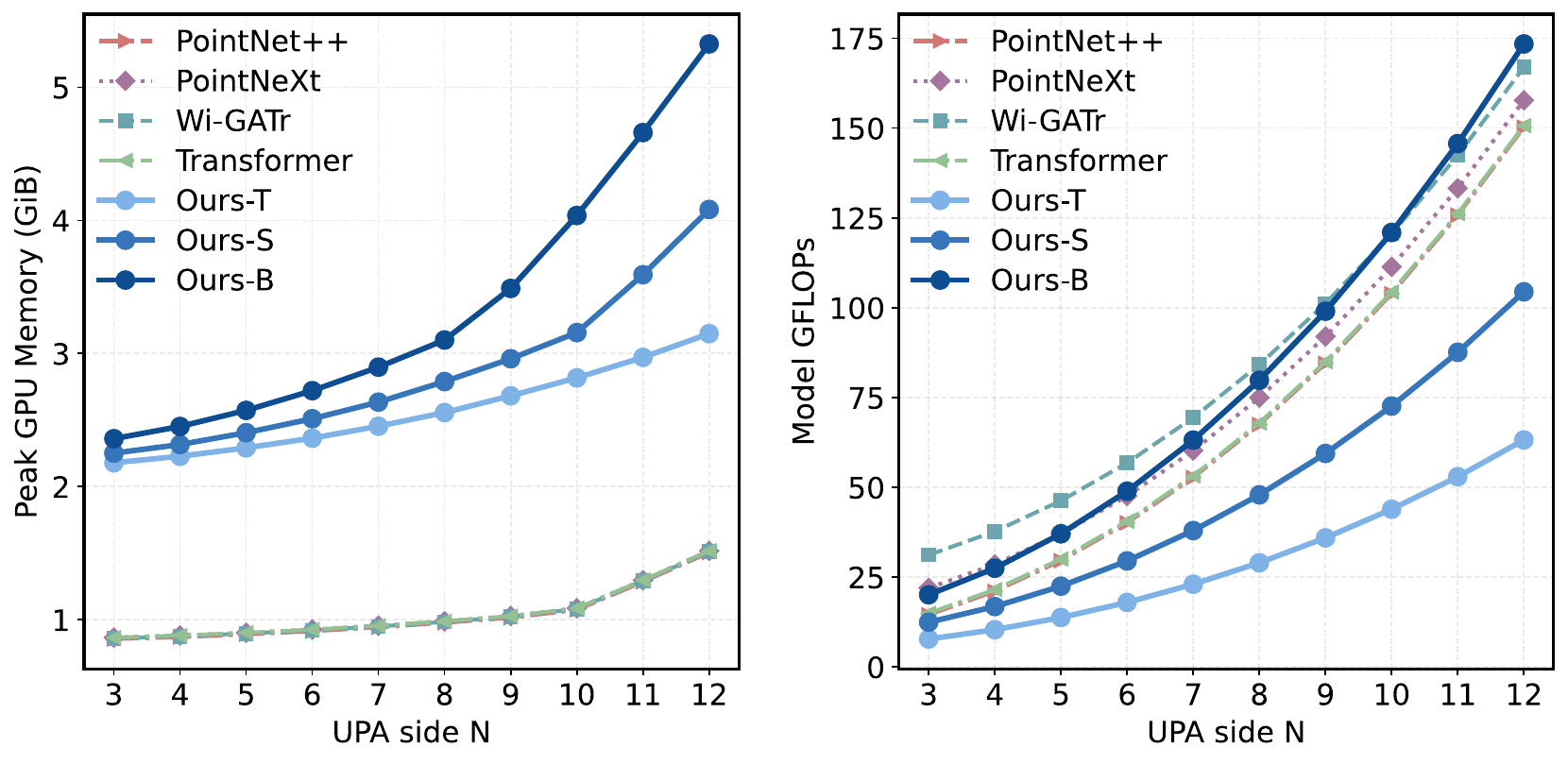}
\caption{\textbf{Computational scaling.} Peak GPU memory is shown on the left and total forward GFLOPs on the right for square UPAs with varying side length $N$.}
\label{fig:array-compute}
\end{figure}

The additional scaling experiments reveal how prediction accuracy and computational requirements change with array size and model capacity. For received power prediction in \cref{fig:array-power-mae}, RFWeave-S and RFWeave-B maintain substantially lower MAE than the feedforward baselines throughout the training range of $N=3,\ldots,8$, with RFWeave-B remaining around 2.2--2.6~dB. At the first unseen array size ($N=9$), Wi-GATr and the RFWeave variants show a clear increase in received power MAE. The corresponding degradation in spatial spectrum SSIM is much milder. This suggests that received power prediction is more sensitive to array size extrapolation, while the spatial structure of the spectrum remains relatively stable. The computational results in \cref{fig:array-compute} show that RFWeave-T and RFWeave-S require substantially fewer GFLOPs than the feedforward baselines across the evaluated array sizes, while the computational cost of RFWeave-B remains comparable to several baselines. Peak GPU memory increases with array size for all models. The increase becomes more pronounced for the larger RFWeave variants, with RFWeave-B requiring the most memory at large array sizes. The accuracy--computation comparison in \cref{fig:model-capacity} further shows that RFWeave-T already achieves an SSIM above 0.83 and a received power MAE of approximately 2.2~dB at only about 10~GFLOPs. It outperforms all feedforward baselines on both metrics while requiring less computation than most of them. RFWeave-S and RFWeave-B further improve both metrics, although the gains become smaller as model capacity increases. Overall, the three variants provide different accuracy--efficiency trade-offs. RFWeave-T achieves strong accuracy at relatively low computational cost, whereas the larger variants offer further improvements at higher computation and memory cost.

\begin{figure}[t]
\centering
\includegraphics[width=0.98\linewidth]{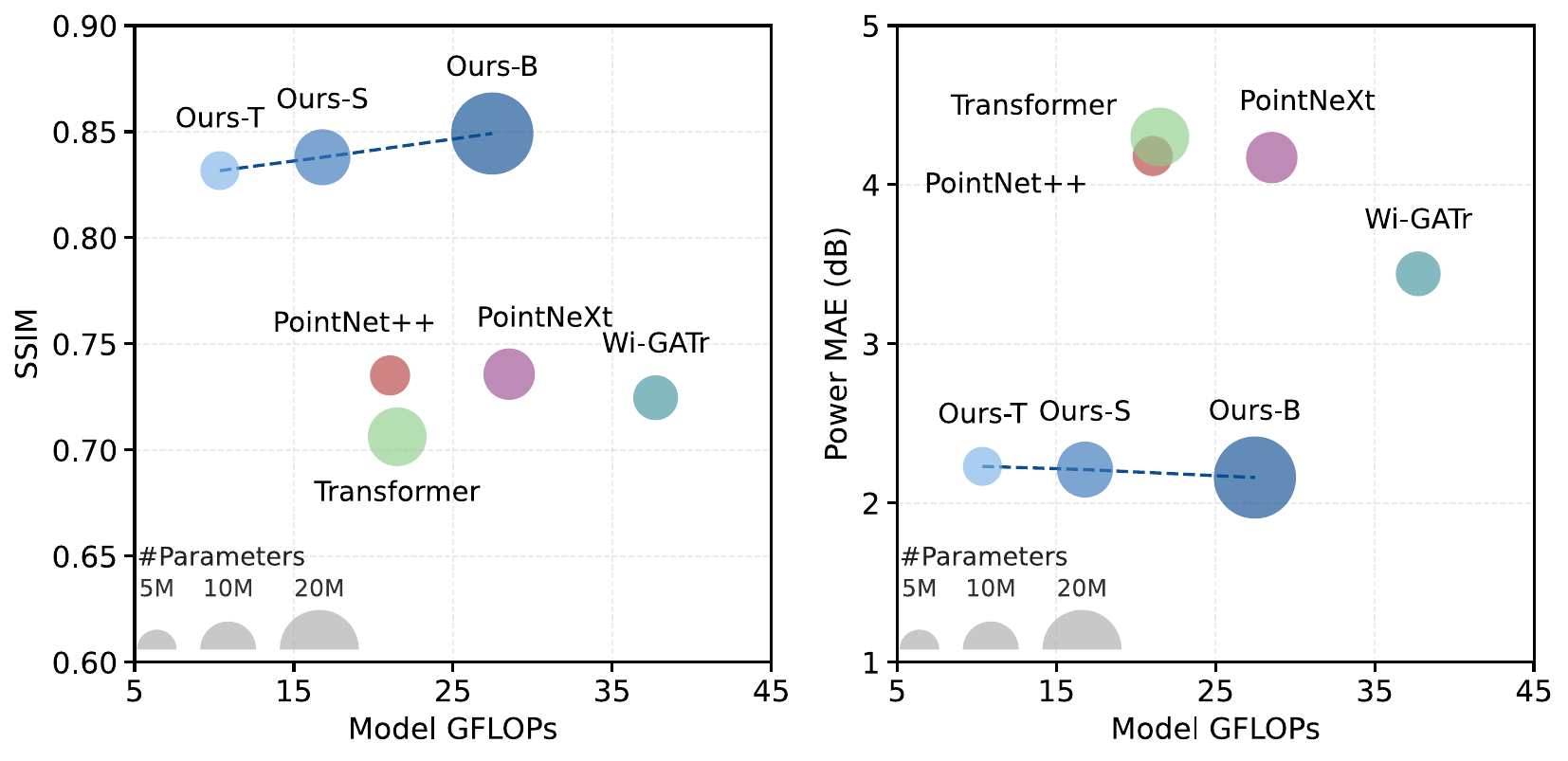}
\caption{\textbf{Accuracy--efficiency trade-off.} Spatial spectrum SSIM is shown on the left and received power MAE on the right against total forward GFLOPs. Bubble size indicates the number of model parameters.}
\label{fig:model-capacity}
\end{figure}

\begin{table}[t]
\caption{Sensitivity to the configuration variation loss weights.}
\label{tab:variance-loss}
\centering
\RFtablefont
\RFtablespacing
\setlength{\tabcolsep}{0.9pt}
\begin{tabularx}{\linewidth}{@{}cc*{6}{Y}@{}}
\toprule
\multirow[c]{2}{*}[-0.20ex]{$\lambda_{\mathrm{rx}}$} & \multirow[c]{2}{*}[-0.20ex]{$\lambda_{\mathrm{beam}}$}
& \multicolumn{3}{c}{Unseen RX Orientation} & \multicolumn{3}{c}{Unseen TX Beam}\\
\cmidrule(lr){3-5}\cmidrule(lr){6-8}
& & \RFhSSIM & \RFwidehSpec & \RFwidehPwr
& \RFhSSIM & \RFwidehSpec & \RFwidehPwr\\
\midrule
0   & 0   & 0.7128 & 3.809 & 2.692 & 0.7205 & 4.051 & 6.789\\
1   & 0   & 0.7485 & 3.525 & 2.771 & 0.7260 & 3.965 & 6.622\\
2   & 0   & 0.7471 & 3.513 & 2.785 & 0.7144 & 4.094 & 6.911\\
0   & 1   & 0.7138 & 3.876 & 3.052 & \RFbest{0.7472} & 4.045 & 7.197\\
0   & 2   & 0.7171 & 3.812 & 2.879 & 0.7349 & 3.955 & 6.845\\
0.5 & 0.5 & 0.6880 & 4.103 & 3.081 & 0.7230 & 4.096 & 6.911\\
\RFhighlight\textbf{1} & \textbf{1} & \RFbest{0.7533} & \RFbest{3.486} & \RFbest{2.253} & 0.7402 & \RFbest{3.870} & \RFbest{6.562}\\
1.5 & 1.5 & 0.7431 & 3.600 & 2.798 & 0.7301 & 4.042 & 6.812\\
2   & 2   & 0.7466 & 3.538 & 2.989 & 0.7276 & 4.079 & 6.850\\
\bottomrule
\end{tabularx}
\end{table}

\subsection{Configuration Variation Loss}
\label{app:variance-loss}

To examine the weight setting of the configuration variation loss, we conduct an ablation using 2,000 WiPTR training rooms with a fixed training budget of 20k steps. Each setting is evaluated on 100 unseen rooms, where we test generalization to unseen receiver orientations and unseen transmit beam directions. We vary the weights of the two variation terms while keeping all other training settings unchanged. \cref{tab:variance-loss} reports the full sweep. The receiver orientation term mainly improves generalization across receiver orientations, while the transmit beam term mainly benefits unseen beam directions. The balanced setting $(\lambda_{\mathrm{rx}},\lambda_{\mathrm{beam}})=(1,1)$ achieves the best performance on five of the six reported metrics and provides the strongest overall results. We therefore use this setting in the main experiments.

\section{Downstream Beam Management}
\label{app:beam-selection}

\paragraph{Evaluation setup.}
We further evaluate whether RFWeave can support downstream beam decisions beyond propagation prediction itself \citep{rezaie2022deviceagnostic,charan2022challenge,klautau2019lidar,alrabeiah2020sub6}. We use the simulated factory environment from \citet{bian2026gsbf}. Both the transmitter and receiver employ a $4\times4$ UPA, and beam selection is performed over a 16-beam DFT codebook. RFWeave-B is initialized from the pretrained checkpoint and then adapted to the factory environment. The factory scene contains five material classes: wood, concrete, metal, chipboard, and marble. These materials are represented using the same continuous material encoding and dataset-specific normalization adopted by the scene tokenizer.

\para{Beam-pair scoring.}
Let $\{\boldsymbol{\omega}_q\}_{q=1}^{Q}$ denote the AoA steering directions on the angular evaluation grid, where $q$ indexes the individual directions. For each transmit beam index $b_t$, RFWeave predicts a peak-normalized AoA spatial spectrum $\widehat S_{\mathrm{dB}}(b_t,\boldsymbol{\omega}_q)$ over these directions, together with a received power estimate $\widehat P_{\mathrm{dB}}(b_t)$. We convert the predicted spectrum from dB to normalized linear spectrum weights,
\begin{equation}
\widehat s(b_t,q)
=
\frac{10^{\widehat S_{\mathrm{dB}}(b_t,\boldsymbol{\omega}_q)/10}}
{\sum_{q'=1}^{Q}10^{\widehat S_{\mathrm{dB}}(b_t,\boldsymbol{\omega}_{q'})/10}},
\qquad
\sum_{q=1}^{Q}\widehat s(b_t,q)=1.
\label{eq:beam-angular-weights}
\end{equation}
Let $\mathbf a_r(\boldsymbol{\omega}_q)$ denote the normalized RX steering vector for direction $\boldsymbol{\omega}_q$, and let $\mathbf w_r^{(b_r)}$ denote the RX codebook vector for beam index $b_r$. We define a score for each TX/RX beam pair as
\begin{equation}
\widehat G(b_t,b_r)
=
10^{\widehat P_{\mathrm{dB}}(b_t)/10}
\sum_{q=1}^{Q}
\widehat s(b_t,q)
\left|
\mathbf a_r(\boldsymbol{\omega}_q)^{\mathsf H}
\mathbf w_r^{(b_r)}
\right|^2.
\label{eq:beam-pair-score}
\end{equation}
This spectrum-based projection combines the relative directional structure encoded by the predicted spatial spectrum with the received power estimate to form a beam-pair ranking score. The resulting score serves to rank candidate beam pairs rather than to directly estimate physical beam-pair power.

Ground-truth beam-pair power is computed directly from the multipath record. Let $g_l$ denote the power of path $l$, and let $\Omega_l^{\mathrm t}$ and $\Omega_l^{\mathrm r}$ denote its departure and arrival directions. Using the TX and RX codebook vectors $\mathbf w_t^{(b_t)}$ and $\mathbf w_r^{(b_r)}$, respectively, the ground-truth power is
\begin{equation}
P_{\mathrm{gt}}(b_t,b_r)
=
\sum_l g_l
\left|
\mathbf w_t^{(b_t)\mathsf H}
\mathbf a_t(\Omega_l^{\mathrm t})
\right|^2
\left|
\mathbf w_r^{(b_r)\mathsf H}
\mathbf a_r(\Omega_l^{\mathrm r})
\right|^2.
\label{eq:beam-gt-power}
\end{equation}
The exhaustive sweep evaluates all beam pairs and selects the pair with the largest $P_{\mathrm{gt}}$, providing an oracle reference.

\para{Top-$K$ selection and spectral efficiency.}
RFWeave ranks all TX/RX beam pairs according to $\widehat G(b_t,b_r)$. Let $\mathcal B_K$ contain the $K$ highest-ranked pairs. To evaluate beam ranking independently of received power calibration, the predicted scores are used only to rank the candidates. Achieved spectral efficiency is then computed using the ground-truth power of the retained beam pairs,
\begin{equation}
R^{(K)}
=
\max_{(b_t,b_r)\in\mathcal B_K}
\log_2\!\left(
1+\frac{P_{\mathrm{gt}}(b_t,b_r)}{P_{\mathrm n}}
\right).
\label{eq:beam-achievable-rate}
\end{equation}
Here, $P_{\mathrm n}$ denotes the noise power, computed from $N_0=-174$~dBm/Hz over a bandwidth of $BW=100$~MHz and converted to linear units. For Top-1, RFWeave retains only its highest-ranked beam pair. Top-2 and Top-3 retain the two and three highest-ranked candidates, respectively. The best spectral efficiency among the retained pairs is then evaluated using ground-truth beam-pair power. Location-based steering produces a single beam pair and is therefore reported only for Top-1. The exhaustive sweep evaluates all beam pairs and provides the oracle upper bound. \cref{tab:beam-management} reports the mean and median spectral efficiency under this protocol.
\endgroup

\end{document}